\documentclass[fleqn,usenatbib]{mnras}

\usepackage{newtxtext,newtxmath}

\usepackage[T1]{fontenc}

\DeclareRobustCommand{\VAN}[3]{#2}
\let\VANthebibliography\thebibliography
\def\thebibliography{\DeclareRobustCommand{\VAN}[3]{##3}\VANthebibliography}

\usepackage{graphicx}	
\usepackage{amsmath}	
\usepackage{xspace}
\usepackage{xcolor}
\usepackage{tabularx}
\makeatletter 
  \patchcmd{\NAT@citex}
    {\@citea\NAT@hyper@{%
      \NAT@nmfmt{\NAT@nm}%
      \hyper@natlinkbreak{\NAT@aysep\NAT@spacechar}{\@citeb\@extra@b@citeb}%
      \NAT@date}}
    {\@citea\NAT@nmfmt{\NAT@nm}%
    \NAT@aysep\NAT@spacechar\NAT@hyper@{\NAT@date}}{}{}

  \patchcmd{\NAT@citex}
    {\@citea\NAT@hyper@{%
      \NAT@nmfmt{\NAT@nm}%
      \hyper@natlinkbreak{\NAT@spacechar\NAT@@open\if*#1*\else#1\NAT@spacechar\fi}%
        {\@citeb\@extra@b@citeb}%
      \NAT@date}}
    {\@citea\NAT@nmfmt{\NAT@nm}%
    \NAT@spacechar\NAT@@open\if*#1*\else#1\NAT@spacechar\fi\NAT@hyper@{\NAT@date}}
    {}{}
\makeatother

\newcommand{\cm}{\,{\rm cm}}

\newcommand{\K}{\,{\rm K}}

\newcommand{\pkpc}{\,{\rm pkpc}}

\newcommand{\LCDM}{\,\rm \Lambda CDM}

\usepackage{enumitem} 

\newcommand{\Eq}[1]{Eq.~(\ref{eq:#1})}

\newcommand{\se}[1]{Section \ref{sec:#1}}

\newcommand{\Fig}[1]{Fig.~\ref{fig:#1}}

\newcommand{\be}{\begin{equation}}
\newcommand{\ee}{\end{equation}}
\newcommand{\bad}{\begin{equation} \begin{aligned}}
\newcommand{\ead}{\end{aligned} \end{equation}}

\newcommand{\Msun}{M_\odot}
\newcommand{\Mpc}{\,{\rm Mpc}}
\newcommand{\kpc}{\,{\rm kpc}}
\newcommand{\pc}{\,{\rm pc}}

\newcommand{\deltac}{\delta_{\rm c}}
\newcommand{\Mv}{M_{\rm vir}}

\newcommand{\Mc}{M_{\rm c}}

\newcommand{\Rv}{R_{\rm vir}}

\newcommand{\Reff}{R_{\rm eff}}

\newcommand{\xihh}{\xi_{\rm hh}}
\newcommand{\ximm}{\xi_{\rm mm}}

\definecolor{mycolor1}{HTML}{54278f}   
\definecolor{mycolor2}{HTML}{756bb1}   
\definecolor{mycolor3}{HTML}{9e9ac8}   
\definecolor{mycolor4}{HTML}{cbc9e2}   

\definecolor{mycolor5}{HTML}{7f2704}   
\definecolor{mycolor6}{HTML}{a63603}   
\definecolor{mycolor7}{HTML}{e6550d}   
\definecolor{mycolor8}{HTML}{fd8d3c}   
\definecolor{mycolor9}{HTML}{fdae6b}   
\definecolor{mycolor10}{HTML}{fee6ce}  

\def\aap{A\&A}

\def\apjl{ApJ}

\def\apjs{ApJS}

\title[Halo assembly bias at high-z]{Halo assembly bias in the early Universe: a clustering probe of the origin of the Little Red Dots}

\author[Z. Wang et al.]{\parbox{17.5cm}{
Zihao Wang$^{1}$,
Fangzhou Jiang$^{2}$\thanks{Corresponding author: fangzhou.jiang@pku.edu.cn},
Haonan Zheng$^{2}$\thanks{hnzheng@pku.edu.cn},
Xuejian Shen$^{3}$,
Zixiang Jia$^{1}$,
Luis C. Ho$^{2}$,
Kohei Inayoshi$^{2}$,
Linhua Jiang$^{2}$
}
\\ \vspace{0.2cm} \\
$^1$ Department of Astronomy, Peking University, Beijing 100871, China \\
$^2$ Kavli Institute for Astronomy and Astrophysics, Peking University, Beijing 100871, China \\
$^3$ Kavli Institute for Astrophysics and Space Research, Massachusetts Institute of Technology, Cambridge, MA 02139, USA \\
\\
}

\date{Accepted XXX. Received YYY; in original form ZZZ}

\pubyear{2026}

\begin{document}
\label{firstpage}
\pagerange{\pageref{firstpage}--\pageref{lastpage}}
\maketitle

\begin{abstract}
The clustering of galaxies encodes key information about the structure and assembly history of their host dark matter (DM) haloes, providing a powerful probe of the origin of extreme high-redshift systems. 
While halo assembly bias has been extensively studied at low redshift, its behavior in the early Universe remains poorly explored.  
Using the large-volume, high-resolution Shin-Uchuu cosmological $N$-body simulation, we characterize halo assembly bias associated with formation time, concentration, and angular momentum across a wide range of halo masses and redshifts.
We find that the sign and amplitude of assembly bias depend on halo mass for both concentration and spin. 
High-concentration and low-spin haloes are more strongly clustered below characteristic peak heights of $\nu \sim 1.5$ and $\sim 0.75$, respectively, while the trends weaken or reverse at higher masses.  
Halo age bias persists at all redshifts but decreases toward higher masses and earlier cosmic times.
We apply these results to assess whether clustering can distinguish competing formation scenarios for the Little Red Dots (LRDs). 
We find that the direct-collapse-black-hole (DCBH) scenario predicts the strongest large-scale bias and enhanced pair fractions, the self-interacting-dark-matter (SIDM) core-collapse scenario and low-spin compact-galaxy scenarios yield weaker clustering due to lower characteristic halo masses and spin-related secondary bias, and the primordial-black-hole (PBH) scenario predicts unbiased clustering.
Our results demonstrate that halo assembly bias and characteristic host masses provide powerful diagnostics for constraining the physical origin of LRDs, offering testable predictions for upcoming clustering measurements with \textit{JWST} and future deep surveys.
\end{abstract}




\section{Introduction}
In the standard $\Lambda$CDM paradigm, dark matter (DM) governs the formation of the large-scale structure (LSS) of the Universe and forms gravitationally bound haloes that provide the sites for galaxy formation and evolution \citep{Press1974, White1991}. 
Galaxies and their host halo trace the otherwise invisible DM distribution;
however, they do not constitute random sampling of the underlying smooth matter density field.
Instead, haloes are clustered and preferentially reside in overdense regions.
This effect is quantified by the {\it halo bias} parameter $b$, defined through 
\be\label{eq:bias}
\xihh(r|\Mv,z) = b^2 \ximm(r,z),
\ee
where $\xihh(r|\Mv,z)$ denotes the two-point correlation function of haloes of virial mass $\Mv$ at redshift $z$ separated by distance $r$, and $\ximm(r,z)$ is that of the underlying DM field.
Halo bias depends primarily on halo mass, with more massive haloes forming preferentially in denser environments \citep[e.g.,][]{Kaiser1984, Sheth1999}.
At fixed mass, however, the clustering strength exhibits a systematic dependence on secondary halo properties, such as formation time and internal structures \citep[e.g.,][]{Gao2005,Gao2007,Wechsler2006,Li2008,Salcedo2018}. 
This dependence of halo clustering on properties beyond mass is referred to as {\it assembly bias}. 

A number of secondary halo properties are known to influence the halo bias parameter, as revealed by cosmological $N$-body simulations. 
Notably, at fixed mass, haloes that form earlier are more strongly clustered than those that form later \citep{Gao2005}. 
The amplitude of this effect is most pronounced for low-mass haloes and progressively weakens towards higher masses \citep[e.g.,][]{Gao2007,Salcedo2018,Sato-Polito2019}.
Relatedly, assembly bias effects are found for halo internal structure: at fixed mass, higher-concentration haloes exhibit stronger clustering \citep[e.g.,][]{Wechsler2006,Gao2007,Salcedo2018} below a characteristic mass scale of $\Mv \simeq 10^{13}\Msun$ at $z=0$. 
Interestingly, this trend reverses at higher masses \citep{Sato-Polito2019}.  
A similar phenomenon has been identified for halo angular momentum, with higher-spin haloes being more strongly clustered, particularly above $\Mv \simeq 10^{12}\Msun$ \citep[e.g.,][]{Bett2007,Faltenbacher2009,Lacerna2012}.
These results demonstrate that the strength of secondary bias effects depends sensitively on halo mass.
It is therefore natural to expect that their magnitude, and potentially even their qualitative behavior, may evolve with cosmic time.
However, this redshift dependence remains relatively unexplored.

Assembly bias provides valuable insight into the formation pathways of galaxies with extreme morphologies. 
For example, \citet{Zhang2025} showed that isolated ultra-diffuse galaxies (UDGs) exhibit surprisingly stronger clustering than that expected for dwarf-mass haloes, which are commonly assumed to host such systems.
This finding has motivated scenarios in which UDGs preferentially reside in the earliest-forming haloes composed of self-interacting dark matter, allowing sufficient time for a shallow DM core to develop.

The \textit{James Webb Space Telescope} (\textit{JWST}) has revealed a diverse population of high-redshift galaxies with morphologies markedly distinct from those in the nearby Universe, ranging from compact and clumpy systems \citep[e.g.,][]{Tanaka2024AGN,Mowla2024,Fujimoto2025,Vanzella2026} to extremely extended disks \citep[e.g.,][]{Wang2025,Umehata2025}.
Their formation, taking the early giant disks as an example, has been hypothesized to be linked to particular halo structures and environments \citep{Jiang2025-BW}.
At the most compact end of this morphological diversity lie the so-called Little Red Dots (LRDs), which have attracted intense attention \citep[e.g.,][]{Matthee2024,Greene2024,Furtak2024,Labbe2023,Kocevski2025}.
LRDs have been proposed as a potentially new class of active galactic nuclei (AGN) host galaxies. They are characterized by a distinctive V-shaped spectral energy distribution and broad Balmer emission lines~\citep[e.g.,][]{Greene2024,Kocevski2025,Hainline2024}, yet lack several canonical AGN signatures, such as strong X-ray emission, prominent near-to-mid-infrared excess, or powerful radio jets \citep[e.g.,][]{Maiolino2025,Yue2024,Akins2024,Perezgonzalez2024,Gloudemans2025}.
Based on these observed signatures, LRDs are inferred to host supermassive black holes (SMBHs) with masses of $\sim 10^6-10^{8}\,\Msun$~\citep[][]{Matthee2024,Greene2024,Taylor2025}, while exhibiting a striking deficiency of stellar components~\citep{Chen2025}. 
The ubiquity of such systems at $z\ga 5$ presents significant challenges to conventional baryonic seeding scenarios and to the standard framework of galaxy-black hole coevolution established in the lower-redshift Universe~\citep[e.g.,][]{Kormendy2013,Reines2016}.

Although the physical nature and the internal structure of LRDs remain under debate~\citep[e.g.,][]{Naidu2025BHS,Lin2025LRD,Inayoshi2025,Kido2025}, a number of scenarios have been proposed to explain their cosmological statistical origin, particularly in terms of the properties required of their host DM haloes. 
First, within the standard $\LCDM$ paradigm, the direct-collapse-black-hole (DCBH) scenario posits that LRDs originate from massive seeds formed in chemically pristine haloes with sufficiently deep potential wells, exposed to strong Lyman-Werner (LW) radiation that suppresses molecular hydrogen cooling and fragmentation \citep[e.g.,][]{Bromm2003,Begelman2006,Lodato2006}. 
While this mechanism provides a viable pathway for forming massive black hole seeds, its ability to reproduce the observed abundance and mass distribution of LRDs remains to be fully quantified in a cosmological statistical framework.

Second, in alternative cosmologies where DM possesses significant self-interactions, LRDs may arise from gravothermal core collapse in self-interacting dark-matter (SIDM) haloes \citep{Jiang2025,Shen2025,Feng2025}.
In this picture, early formation and high halo concentration are key conditions required to trigger sufficiently rapid core collapse and SMBH seeding by the observed epoch \citep{Jiang2025}.

Beyond scenarios centered on massive BH seed formation, additional interpretations do not necessarily explicitly require LRDs to host SMBHs. 
For example, \citet{Pacucci2025} argue that, based on the observed abundance, compactness, and redshift distribution of LRDs, these systems may preferentially inhabit haloes in the extreme low-spin tail of the angular momentum distribution. 
In this model, LRDs emerge as compact galaxies lacking rotational support due to the intrinsically low spin of their host haloes \citep{Mo1998}. 

Although many of the aforementioned scenarios can reproduce halo number densities broadly consistent with the observed abundance of LRDs, they imply markedly different host halo properties, such as formation time, concentration, or spin.
The clustering strength of their host haloes, therefore, may provide a powerful and complementary diagnostic of LRD origins, particularly in light of forthcoming large-scale clustering measurements from \textit{JWST} surveys.
However, while assembly bias has been extensively studied in cosmological simulations at low redshift \citep[$z \lesssim 3$; e.g.,][]{Contreras2019}, its behavior in the early Universe ($z \gtrsim 5$) remains poorly understood. 
With the advent of $N$-body simulations that simultaneously achieve high mass resolution and large cosmological volume, it is timely to revisit halo assembly bias at high redshift and explore how halo clustering can be used to constrain the formation pathways of LRDs.

In this work, we extend measurements of halo assembly bias associated with halo age, concentration, and spin -- the  properties that are closely related to proposed formation pathways of LRDs -- to redshifts as high as $z=10$.  
Using the high-resolution dark-matter-only (DMO) simulation Shin-Uchuu~\citep{Ishiyama2021}, we quantify how these bias signals evolve with mass and redshift, providing an important reference for future clustering analyses. 
We then apply these results, for the first time, to LRD formation scenarios by populating haloes with LRDs according to the aforementioned models.

This paper is organized as follows. 
In \se{sim}, we introduce the simulations used in this work and list the definitions of relevant halo properties. 
In \se{halobias}, we describe the measurements of halo assembly bias signals and examine their redshift evolution. 
In \se{lrdbias}, we outline how LRD host haloes are selected in the simulation under different physical scenarios, compare their cross-correlations with those of normal galaxies, and discuss the implications for LRD formation pathways. 
We summarize the main conclusions in \se{conclusions}. 
Throughout the paper, we adopt the cosmology used in the Shin-Uchuu simulation, and define haloes as spherical overdensities that are 200 times the critical density of the Universe.

\section{Simulation and methods}\label{sec:sim}

\subsection{Simulation}
\begin{figure}
    \centering
    \includegraphics[width=0.95\linewidth]{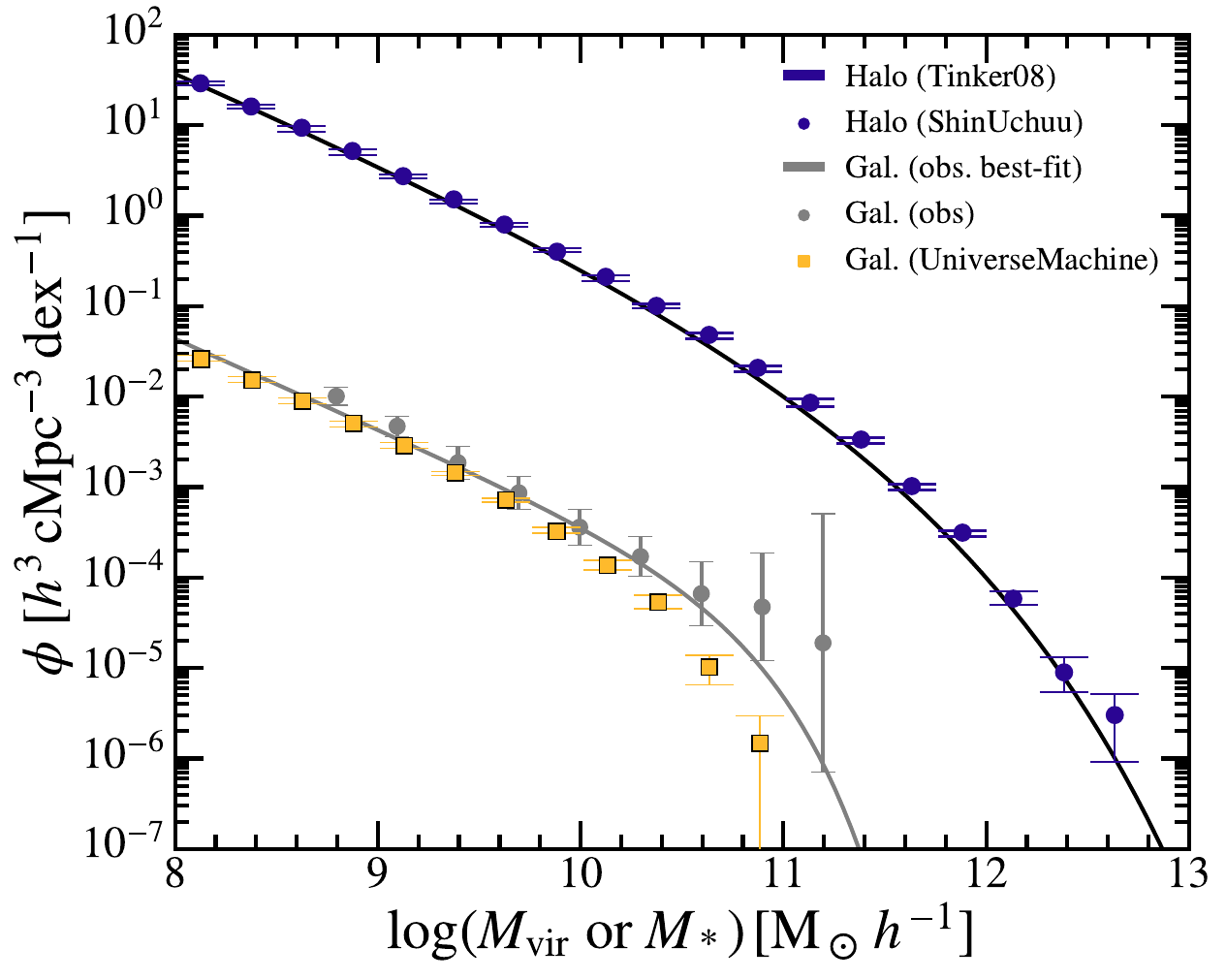}
    \caption{Halo and stellar mass functions at $z\simeq5$. Blue circles show the distinct halo mass function from Shin-Uchuu as a function of halo virial mass $M_{\rm vir}$, while yellow squares show the stellar mass function from the UniverseMachine catalogue as a function of stellar mass $M_*$. The halo mass function is in good agreement with theoretical expectations from ~\citealt{Tinker2008} (blue solid), while the stellar mass function exhibits good agreement at the low-mass end and a deficit at the massive end relative to recent \textit{JWST} results from \citealt{Wang2025MIRI} (gray).}
    \label{fig:hmf}
\end{figure}

To characterize the clustering signal of the haloes of interest, we utilize the Shin-Uchuu $N$-body simulation~\citep{Ishiyama2021,Oogi2023,Dong2024,Aung2023,Prada2023}. The simulation has a comoving box of side-length $140\, h^{-1}\Mpc$, particle mass of $8.97 \times 10^5\,h^{-1} \Msun$, and a gravitational softening length of $0.4\, h^{-1} \kpc$. DM haloes are identified with the \textsc{rockstar}\footnote{\url{https://bitbucket.org/gfcstanford/rockstar/}} halo finder~\citep{Behroozi2012}, which employs six-dimensional phase-space and temporal information to separate overlapping structures and to track subhaloes across snapshots. 
The halo center corresponds to the position of the particle with the minimum potential.
In this work, we use the group catalogue provided by the Shin-Uchuu team\footnote{\url{https://www.skiesanduniverses.org/Simulations/Uchuu/}}, which includes pre-computed halo properties. Merger trees are constructed with a modified, parallelized implementation of the \textsc{consistent tree}\footnote{\url{https://bitbucket.org/pbehroozi/consistent-trees/}} algorithm~\citep{Behroozi2013}. 
Throughout the analysis, we consider haloes with more than $100$ particles as resolved. A halo is regarded as relaxed if it satisfies $X_{\rm off}<0.07$ and $2T/|U|<1.5$, where $X_{\rm off}\equiv{|\vec{r}_{\rm center}-\vec{r}_{\rm com}|}/{R_{\rm vir}}$ is the offset between the minimum-potential center $\vec{r}_{\rm center}$ and the center of mass $\vec{r}_{\rm com}$, and $2T/|U|$ is the virial ratio, where $U$ and $T$ are the potential and kinetic energies, respectively. 
In Fig.~\ref{fig:hmf}, we show the halo mass function measured from Shin-Uchuu at $z\simeq5$, and compare it with the theoretical prediction from the Press-Schechter theory with the ellipsoidal-collapse critical overdensity \citet{Sheth2001}. 

\subsection{Measurement of halo properties}
We investigate halo assembly bias across several halo properties, focusing in particular on concentration, spin, and formation time. 
Since multiple definitions of these parameters are provided in the halo catalogue, here we specify the definition that we adopt.

We use the concentration obtained by fitting the Navarro-Frenk-White \citep[NFW;][]{Navarro1997} density profile
\begin{equation}
    \rho(r)=\frac{\rho_{\rm s}}{\left(r/r_{\rm s}\right)\left(1+r/r_{\rm s}\right)^{2}},
\end{equation}
where $\rho_{\rm s}$ is the characteristic (scale) density and $r_{\rm s}$ is the scale radius.
Halo particles are divided into up to 50 radial equal-mass bins, with a minimum of 15 particles per bin, and an NFW profile is directly fitted to the density distribution to obtain the maximum-likelihood estimate of $\rho_s$ and $r_{\rm s}$. We then construct concentration as $c = R_{\rm vir}/r_{\rm s}$.
The scale density $\rho_{\rm s}$ is not saved in the catalogue --when needed, we derive it from the relation,
\begin{equation}
\label{eq:rhos}
\rho_{\rm s}=\frac{M_{\rm vir}}{4\pi r_{\rm s}^{3}\left[\ln(1+c)-c/(1+c)\right]}.
\end{equation}

We use the spin parameter in the \citet{Bullock2001} definition, given by 
\begin{equation}
    \lambda =\frac{J_{\rm vir}}{\sqrt2M_{\rm vir}R_{\rm vir}V_{\rm vir}}
\end{equation}
where $J_{\rm vir}$ is the total angular momentum of all bound DM particles within the virial radius, and $V_{\rm vir}$ is its circular velocity at the virial radius.

We define the halo formation epoch as the (earliest) redshift when the halo mass exceeds half its present virial mass, $z_{1/2}$.
The corresponding look-back time of formation is then
\begin{equation}
    t_{\rm form} = t(z) - t(z_{1/2}),
\end{equation} where $t(z)$ is the cosmic time at redshift $z$.

The prominence of a DM halo relative to the cosmic density fluctuations is commonly quantified by the peak height of the associated overdensity, defined as 
\be\label{eq:nu}
\nu \equiv \deltac / \sqrt{S(M_{\rm vir})},
\ee
where $\deltac(z)$ denotes the critical linearly extrapolated overdensity for halo collapse, and $S(M_{\rm vir})$ is the variance of the density field smoothed on the mass scale $\Mv$.

\begin{figure}
    \centering
    \includegraphics[width=0.9\linewidth]{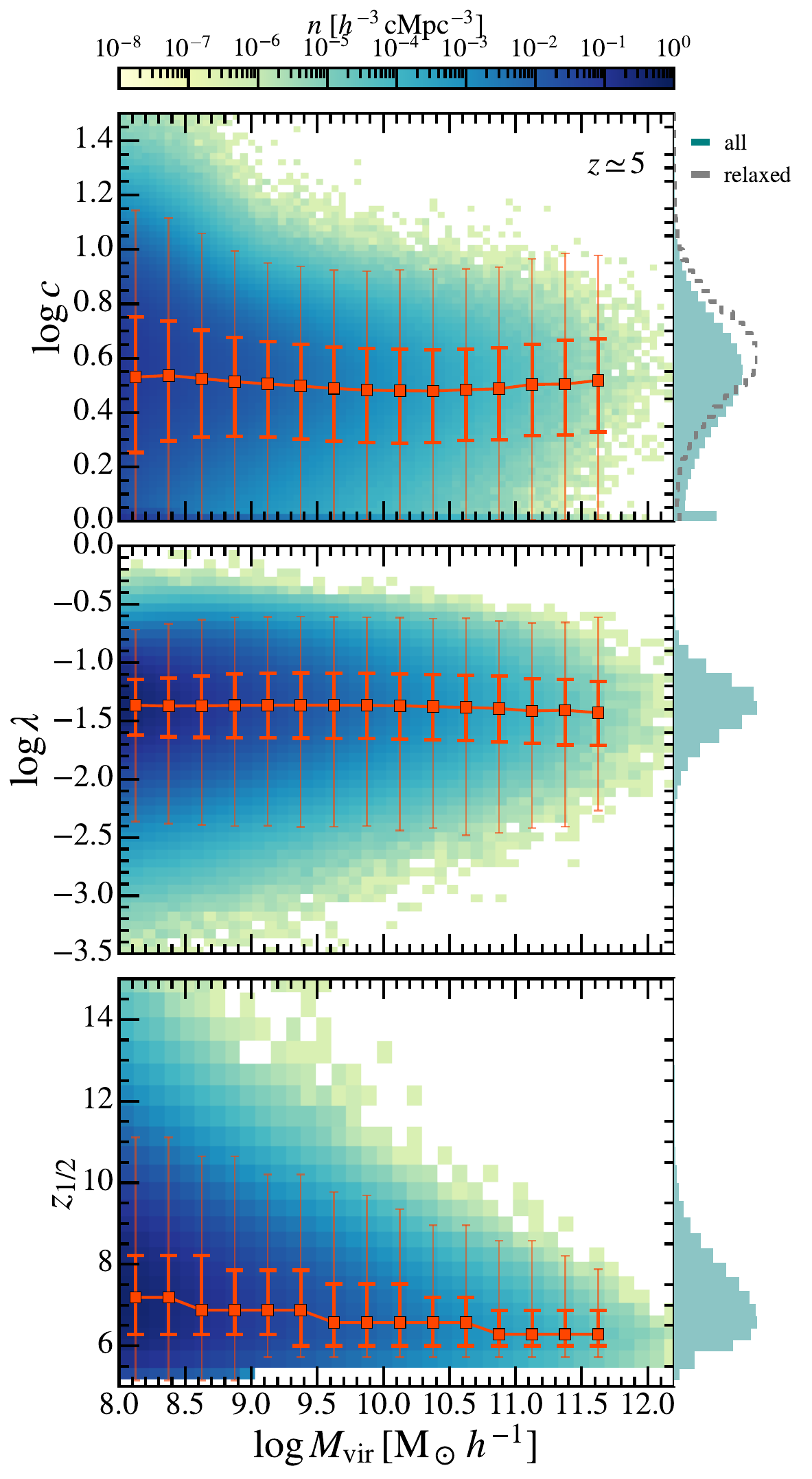}
    \caption{Distributions of halo properties examined in this study, at $z\simeq 5$ in the Shin-Uchuu simulation. 
    From top to bottom, the panels display halo concentration, spin, and formation redshift as functions of halo mass. 
    The colour indicates number density, while the solid curves indicate the median relation. The thicker and thinner error bars denote the $1\sigma$ and $4\sigma$ intervals, respectively. 
    Side panels show the corresponding one-point distributions. 
    For concentration, the histogram of the subsample of relaxed haloes is included for comparison, as indicated by the dashed grey line. 
    At high redshift, halo concentration exhibits no clear dependence on mass. 
    Spin is nearly mass-independent.
    Lower-mass haloes tend to form earlier than more massive ones.}
    \label{fig:scaling}
\end{figure}

Fig.~\ref{fig:scaling} presents the distributions of concentration, spin, and formation time as functions of halo mass for all resolved distinct haloes at $z=5$. 
In contrast to the concentration-mass relation at low redshift~\citep[e.g.,][]{Prada2012,Ludlow2012,Dutton2014,Ishiyama2021}, which exhibits a negative slope, the halo concentration at cosmic dawn is nearly mass-independent, with a median of $\simeq 3$, consistent with recent findings in other cosmological simulations \citep[e.g.,][]{Ishiyama2021,Yung2024}. 
A pronounced low-$c$ tail extending to $c \sim 1$ is evident and is primarily associated with perturbed haloes, for which concentration is ill-defined. 
Excluding these systems shifts the distribution towards higher $c$ values. 
The median spin parameter shows little correlation with mass, with a median value of $\lambda \simeq 0.04$ and no significant redshift evolution, consistent with previous findings \citep[e.g.,][]{Bullock2001}. 
The formation times of low-mass haloes are generally earlier than more massive ones, reflecting the hierarchical nature of structure formation. 

\subsection{UniverseMachine galaxy catalogue}
To model the galaxy population, we use the publicly available UniverseMachine galaxy catalogue \citep{Behroozi2019}. 
UniverseMachine is a forward modeling framework that populates galaxies within cosmological $N$-body simulations by assigning star formation histories, stellar masses, and UV luminosities through an empirically calibrated model of galaxy growth. 
The framework describes star formation as a function of halo mass, assembly history, and redshift using a flexible parameterization, with parameters constrained by jointly fitting a wide range of observational data, including stellar mass functions, UV luminosity functions, the cosmic star formation rate density, quenched fractions, and, in particular, galaxy clustering over the redshift range $z=0$ to $z\simeq10$. 
The explicit calibration to clustering statistics ensures that the spatial distribution of galaxies is consistently modeled alongside their internal properties, thereby providing a robust foundation for subsequent clustering analyses.

In this work, we adopt the $z\simeq5$ mock galaxy catalogue constructed on the Shin-Uchuu halo merger trees. As shown in Fig.~\ref{fig:hmf}, the UniverseMachine stellar mass function at $z\simeq5$ is broadly consistent with recent \textit{JWST} measurements~\citep{Wang2025MIRI}, with only a slight deficit at the massive end.  We use the predicted stellar masses, star formation rates (SFRs), UV magnitudes, and galaxy-halo connections to identify candidate LRD hosts under certain formation scenarios discussed in Sect.~\ref{sec:lrdbias}.

\section{Redshift evolution of halo assembly bias}\label{sec:halobias}

In this section, we investigate how the clustering of haloes depends on concentration, spin, and formation time, and characterize the redshift evolution of the halo assembly bias. 

\subsection{Auto-correlation function}\label{sec:2pcf}
We base our analysis on all resolved distinct haloes, and compute the auto-correlation function (ACF) using
\begin{equation}
    \xi(r)=\frac{DD(r)}{RR(r)}-1
\end{equation}
where $DD(r)$ and $RR(r)$ denote the normalized numbers of halo-halo and random-random pairs, respectively, counted in bins of separation $r$. These quantities are defined as
\bad
DD(r)&=\frac{N_{DD}}{N_D(N_D-1)/2},\\
RR(r)&=\frac{N_{RR}}{N_R(N_R-1)/2},
\ead
where $N_D$ ($N_R$) is the number of haloes (random points); $N_{DD}$ and $N_{RR}$ are the numbers of halo-halo and random-random pairs, respectively.
The calculation is performed for all haloes in the simulation box that satisfy the selection criteria, using the publicly available \textsc{corrfunc} code~\citep{Sinha2020}. 
Haloes are further grouped into bins of width $0.25$ in $\log (M_{\rm vir}/\Msun)$ spanning the range $8$ to $13$.
We require at least 1000 pairs per mass bin to obtain statistically robust measurements.

\begin{figure*}
    \includegraphics[width=0.95\linewidth]{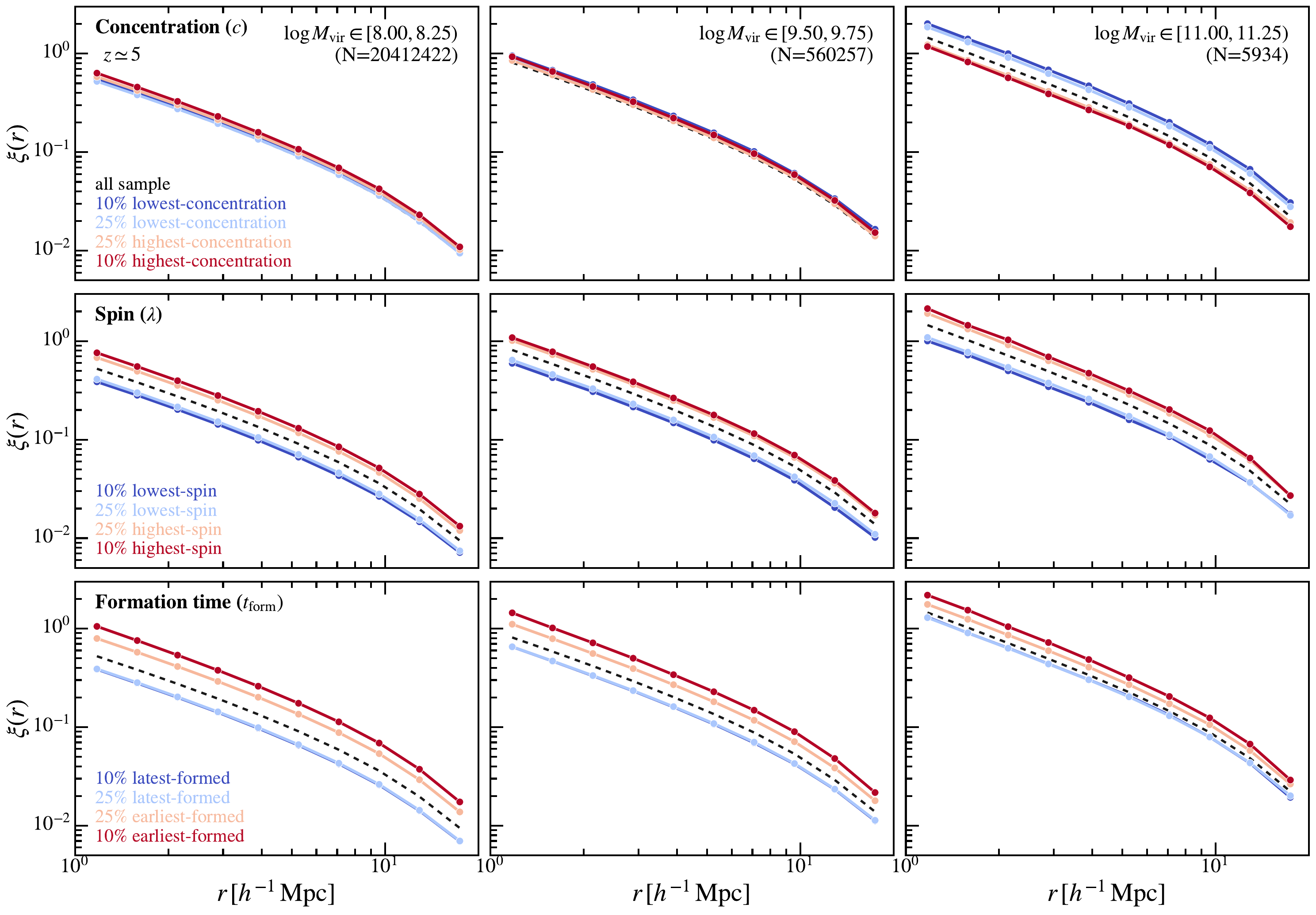}
    \caption{Auto-correlation functions (ACFs) of haloes from the Shin-Uchuu simulation at $z \simeq 5$. 
    Haloes are grouped into mass bins of width 0.25 dex, and we show three representative cases: low mass ($\log( M_{\rm vir}\,[h^{-1}\rm \Msun])\simeq8$), intermediate mass ($\log( M_{\rm vir}\,[h^{-1}\rm \Msun])\simeq9.5$), and high mass ($\log( M_{\rm vir}\,[h^{-1}\rm \Msun])\simeq11$). 
    The dashed black curves denote the ACFs of the full populations in each mass bin, while the colored curves correspond to subsamples selected by percentiles of concentration, spin, or formation time, as indicated. The three parameters exhibit distinct assembly-bias behaviors and mass trends. 
    Concentration shows weak bias at the low-mass end and a reversal at high masses, where low-$c$ haloes cluster more strongly. 
    Spin exhibits a mass-independent trend in which low-spin haloes are more strongly clustered.
    Formation time shows enhanced clustering for the oldest haloes, with the strength of this trend weakening towards higher masses.}
    \label{fig:xi}
\end{figure*}

\Fig{xi} presents the ACFs at $z \simeq 5$ in three halo mass bins, subdivided by percentile ranges of each halo parameter under consideration. 
The results reveal both similarities and differences relative to the trends observed in the local Universe. 

At the low-mass end, $\Mv\simeq 10^8 h^{-1}\Msun$, corresponding to an overdensity peak height of $\nu\simeq 1.3$, 
haloes with higher concentrations are marginally more strongly clustered than their low-concentration counterparts.\footnote{At such high redshift, more than half of the haloes are classified as perturbed under our criteria, with a non-negligible fraction exhibiting $c=1$. In our default analysis, these systems belong to the low-concentration subsamples and could bias the clustering of the low-$c$ population. However, excluding these perturbed haloes yields statistically consistent results, indicating that this effect does not alter our conclusions.} For reference, the characteristic halo mass at this redshift is much lower, $M_\star(\nu=1)\simeq 10^{6.03}\,h^{-1}\Msun$.
This trend exhibits a mass dependence, weakening toward higher masses and eventually reversing: 
by $\Mv\simeq10^{11}h^{-1}\Msun$, haloes in the lowest 10\% of the concentration distribution are nearly twice as strongly clustered as those in the highest 10\%. 
In the local Universe, this nonlinear dependence is often connected to the statistics of density peaks in the initial field~\citep{Dalal2008}. At fixed $\nu$, peaks of lower curvature are more strongly clustered and tend to assemble their mass earlier than high-curvature peaks, suggesting a connection to the concentration dependence of halo bias. At the low-mass end, however, assembly bias is generally attributed to nonlinear environmental effects that may depend on the location of haloes within the cosmic web~\citep[e.g.,][]{Sandvik2007,Desjacques2008,Hahn2009,Borzyszkowski2017}, for which a comprehensive theoretical framework has yet to be established and is beyond the scope of this work.

High-spin haloes exhibit consistently stronger clustering across the mass range probed here, and the strength of this spin-dependent bias shows little variation with halo mass, in contrast to the trend observed at $z=0$~\citep[e.g.,][]{Sato-Polito2019}. In the local Universe, dwarf low-spin haloes are more strongly clustered at low redshift, a trend commonly attributed to splashback haloes~\citep[e.g.,][]{Tucci2021}. These systems reside near massive hosts and inherit the large-scale bias of their neighbors. At $z=5$, however, splashback haloes are expected to be extremely rare. The trend we observe therefore likely reflects the intrinsic spin-dependent bias, which may arise from tidal torques exerted by nearby large-scale structures: because the tidal field that generates halo angular momentum correlates with the surrounding density field, high-spin haloes preferentially reside in more strongly clustered environments \citep{Lacerna2012,Paranjape2018}.

In the bottom row, the oldest haloes are more strongly clustered than the youngest ones, and this age-dependent signal also displays a notable mass dependence consistent with the local relation. A possible interpretation is that older haloes preferentially form in denser environments~\citep{Borzyszkowski2017,Contreras2021}, while at the low-mass end the age-dependent bias is further driven by environmental quenching of halo growth, such as tidal suppression of mass accretion near massive neighbours~\citep{Diemand2007,Mansfield2020}.

\subsection{Redshift evolution of the assembly bias}
\Fig{xi} has demonstrated that the assembly bias signal exhibits a dependence on halo mass. 
We now systematically quantify the mass dependence using the {\it relative bias} factor, defined as the ratio of the ACF of haloes selected by a given secondary property to that of the unconditional halo population at the same mass,
\begin{equation}\label{eq:RelativeBias}
    \tilde{b}_X^2(r|\Mv)=\frac{\xi_X(r|\Mv)}{\xi(r|\Mv)}.
\end{equation}
Here, $\xi_X(r|\Mv)$ denotes the ACF of haloes selected by the secondary property, $X$ (e.g., concentration, spin, or formation time), while $\xi(r|\Mv)$ represents the ACF of the full halo population within the same virial mass bin. 
Unless otherwise specified, the relative bias is evaluated over scales of $5$-$15\,h^{-1}\,\mathrm{Mpc}$ to facilitate  direct comparison with previous studies~\citep[e.g.,][]{Tucci2021,Sato-Polito2019}. 
We estimate uncertainties using $1000$ bootstrap re-samplings with replacement, and find that in most cases the resulting uncertainties are small. 
The bias factors are then computed as the error-weighted mean of $\tilde{b}_X^2(r|\Mv)$ over the adopted scale range, where the uncertainties are estimated via bootstrap re-sampling. 

\begin{figure*}
    \includegraphics[width=0.9\linewidth]{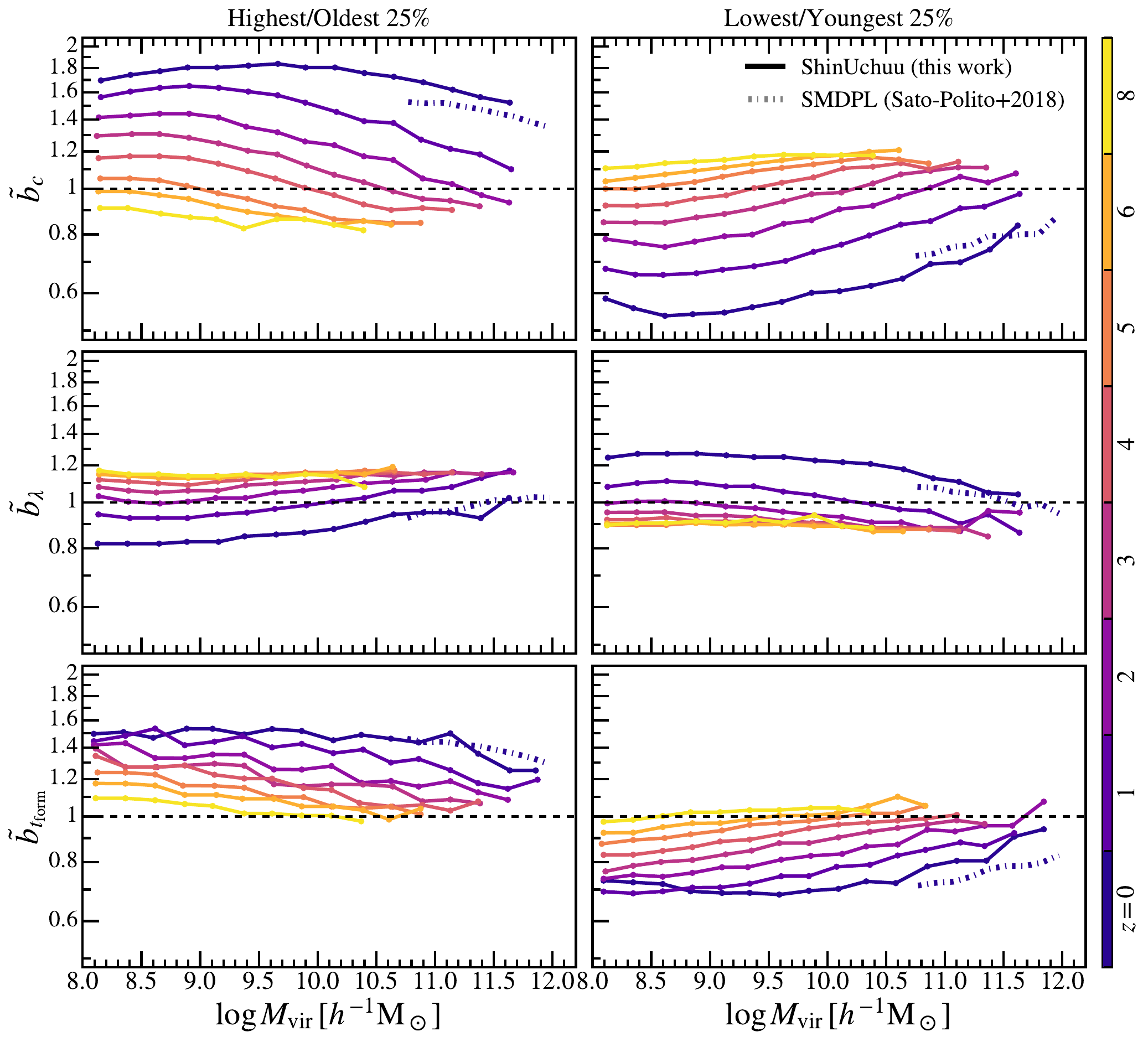}
    \caption{Redshift evolution of secondary bias as a function of halo mass for concentration (top), spin (middle), and formation time (bottom), from $z=0$ to $z\simeq8$. The y-axis is displayed on a logarithmic scale, with tick labels indicating the corresponding values for ease of interpretation. In each panel, the left-hand column shows the bias factors for the upper 25\% of the parameter distribution, and the right-hand column shows the corresponding lower 25\%. Curves are evaluated over scales of $5-15\,h^{-1}\,\Mpc$. At $z=0$, we assess consistency by comparing our measurements with results from \citet{Sato-Polito2019}. The amplitudes of all the secondary biases here exhibit dependence on halo mass and clear evolution with redshift.
}
    \label{fig:bias_z_evo}
\end{figure*}

\begin{figure*}
    \centering
    \includegraphics[width=0.9\linewidth]{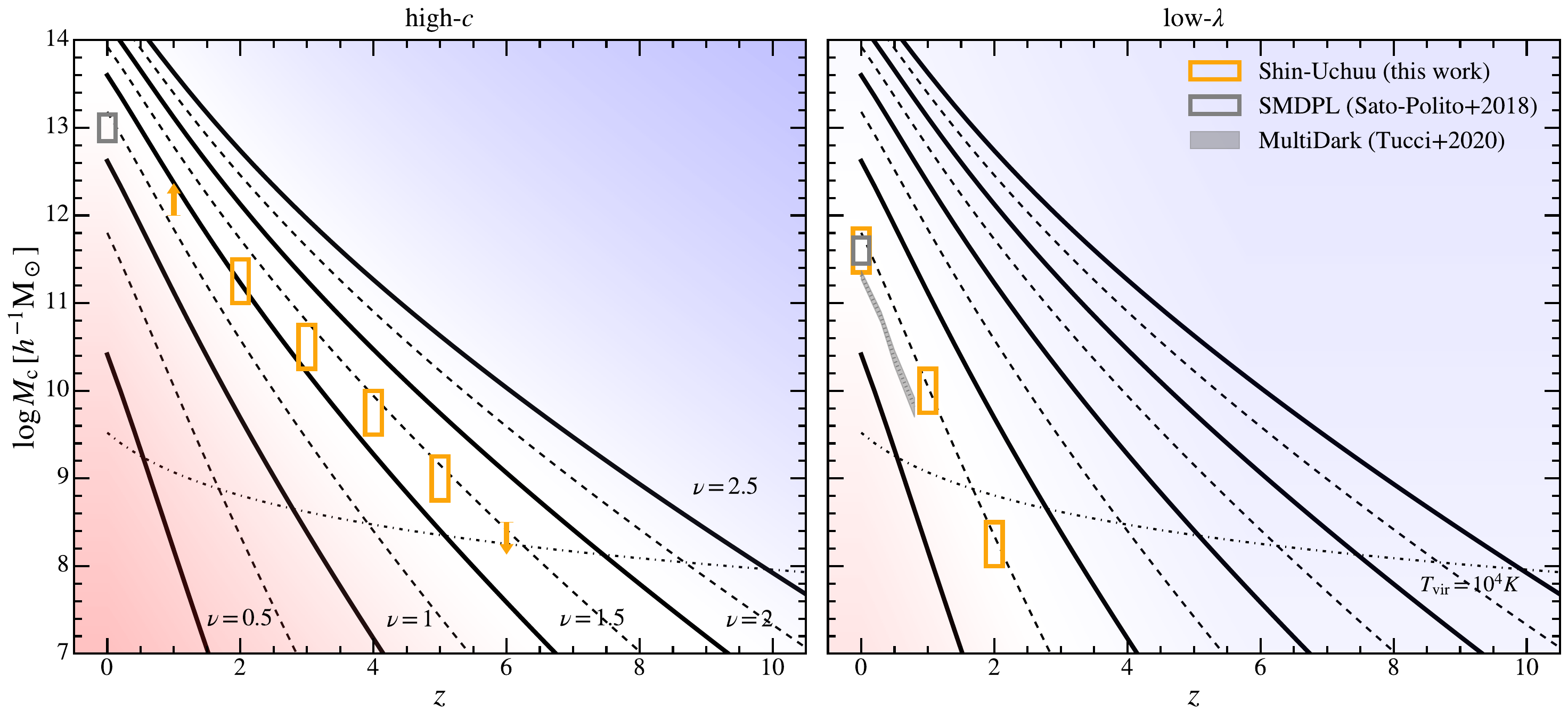}
    \caption{
    Redshift evolution of the critical mass, $\Mc$, at which the assembly-bias signals reverse sign for high-concentration (left) and low-spin (right) haloes. 
    The critical masses are extracted from \Fig{bias_z_evo} as the halo-mass intervals within which the relative bias (\Eq{RelativeBias}) of the top 25\% highest-$c$ or bottom 25\% lowest-$\lambda$ haloes crosses unity. 
    Overlaid for comparison are the contours of peak height ($\nu$), measurements at low redshift from previous studies (grey symbols or band), and the atomic-cooling threshold, which serves as a proxy for the minimum halo mass for galaxy formation. 
    A red (blue) background colour indicates the regimes of more (less) strong clustering, with the colour saturation schematically illustrating the degree of deviating from a relative bias of unity. \quad
    The critical masses for high-$c$ bias and low-$\lambda$ bias broadly correspond to $1.5-1.75\sigma$ and $0.75\sigma$ density peaks, respectively.}
    \label{fig:crossM}
\end{figure*}

\Fig{bias_z_evo} presents the relative bias of the upper and lower quartiles of each parameter as a function of halo mass, evaluated over redshifts $0\le z \le 8$. 
The relative bias associated with halo concentration exhibits a strong dependence on both mass and redshift. 
At $z\gtrsim 6$, low-concentration haloes are marginally more strongly clustered than the full population, while high-concentration haloes are less clustered, across the entire mass range probed. 
As cosmic time progresses, the signal weakens and reverses sign at a redshift-dependent transition mass. 
For example, at $z\simeq5$ (4), the sign flip for the high-$c$ subsample occurs at $\Mv\sim10^{9.0}h^{-1}\Msun$ ($10^{9.8}h^{-1}\Msun$). 
At later times, this crossover mass steadily shifts toward higher values. 
Meanwhile, at fixed halo mass below the transition scale, the bias amplitude increases monotonically with redshift, likely reflecting that halos of a given mass arise from increasingly rarer density peaks at earlier times~\citep[e.g.,][]{Bardeen1986}. For the mass range considered here ($\Mv=10^8-10^{12}\,h^{-1}\Msun$), the corresponding peak heights increase from $\nu\simeq0.29-0.80$ at $z=0$ to $\nu\simeq2.08-5.66$ at $z=8$.
By $z=0$, concentration exhibits the strongest secondary-bias signal among the three parameters considered here.

The relative bias associated with halo spin exhibits marginal dependence on mass or redshift at $z \gtrsim 4$, where low-spin haloes are consistently less strongly clustered than high-spin haloes across all masses. 
Beginning around $z=3$, the clustering of low-spin haloes progressively increases, eventually reversing sign at the low-mass end.
The sign flip of the assembly bias of the low-$\lambda$ subsample occurs at $\Mv \simeq 10^{10.0} h^{-1}\Msun$ ($10^{11.5}h^{-1}\Msun$) at $z\simeq 1$ (0).
As discussed above, this inversion could be attributed to splashback haloes. \citet{Ma2025} further argued that the reversal may also be driven by tidal effects in dense environments, including cosmic filaments, where low-mass haloes can lose high-angular-momentum material from their outer regions, thereby enhancing the clustering of the remaining low-spin haloes. Our results provide further support for these interpretations: the reversal first appears among low-mass haloes and emerges around the epoch when proto-clusters begin to form.

The oldest haloes are more strongly clustered than the youngest ones across nearly all redshifts. The strength of this age bias decreases with increasing halo mass and with increasing redshift.
Overall, all bias signals are rather weak at cosmic dawn.

As a sanity check, our measurements at $z=0$ based on the high-resolution Shin-Uchuu simulation are broadly consistent with previous studies that used larger-volume simulations with coarser resolution \citep[e.g.,][]{Sato-Polito2019}.

Since the assembly-bias signals associated with both concentration and spin exhibit clear crossover behaviors, it is useful to characterize the redshift evolution of the crossover mass scales, $\Mc$. As shown in \Fig{bias_z_evo}, both ends of each parameter distribution exhibit similar crossover behavior, although they are not perfectly symmetric in the sense that $\tilde{b}_{X,\rm highest}\times\tilde{b}_{X,\rm lowest}=1$~\citep[e.g.,][]{Salcedo2018}. For demonstration, we quantify the crossover mass using the high-concentration and low-spin populations that are associated with the formation of LRDs.
We identify the crossover mass as the halo-mass range over which the relative bias crosses unity, using mass bins of a width of 0.5 dex.
For redshifts at which no clear crossing is detected within the probed mass ranges, we indicate the crossover scale as an upper or lower limit. 
The results are shown in \Fig{crossM}. 
These measurements are compared with contours of peak height.

As shown in the left panel of \Fig{crossM}, the crossover mass for the assembly bias associated with concentration broadly corresponds to $1.5\sigma$ density peaks, ranging from $\sim1.25\sigma$ at $z=1$ to $\sim1.75\sigma$ at $z=6$.
Extrapolation toward lower redshift yields values consistent with the results of \citet{Sato-Polito2019} at $z=0$. 

We further compare the crossover mass with the halo mass that corresponds to the atomic-cooling threshold, given by the virial-temperature criteria: 
\begin{equation}
T_{\rm vir}\equiv\frac{\mu m_{\rm p}}{2k_{\rm B}} V_{\rm vir}^{2}> 10^{4}\K,
\end{equation}
where $\mu\simeq0.6$ is the mean molecular weight, $m_{\rm p}$ is the proton mass, and $k_{\rm B}$ is the Boltzmann constant.
This approximately corresponds to a halo mass above which haloes are able to form galaxies. 
This comparison reveals that galaxies residing in high-concentration haloes at $z\gtrsim 6$ tend to be less strongly clustered than the general galaxy population at the same mass scale.

The right panel of \Fig{crossM} presents an analogous analysis for the spin bias, and identifies the crossover masses below which low-spin haloes become more strongly clustered than their high-spin counterparts. 
The crossover masses broadly correspond to $0.75\sigma$ density peaks.
The low-redshift results are in good agreement with previous measurements~\citep[e.g.,][]{Sato-Polito2019,Tucci2021}. 
Comparison with the atomic cooling threshold indicates that at $z \gtrsim 2$, the low-spin haloes that can host galaxies are less clustered than the general population.

\section{Implications on the origin of LRDs}\label{sec:lrdbias}

The halo assembly bias at high redshifts, as characterized above, may provide a useful handle for identifying the host haloes of galaxies with unusual properties. 
For example, the various proposed scenarios for the LRDs attribute their origin to distinctive host-halo properties and therefore may predict distinct clustering signatures.
In this section, we quantify these signatures. 
We emphasize that the discussion here focuses on where LRDs preferentially emerge in terms of their host halo properties, while the physical nature of LRDs themselves lies beyond the scope of this work. 

We consider four representative scenarios, each imposing distinct constraints on the properties of their host DM haloes or environments. 
We briefly recap these scenarios and present the corresponding prescriptions for assigning LRDs to host haloes in \se{mech}. 
We then analyze the resulting clustering signals on large and small scales in \se{ls_clstr}. 
The uncertainties associated with the tagging methods are discussed in \se{limit}. 

Before diving into specific models, we summarize some general statistics for reference: the LRD number density remains nearly constant from $z\simeq4$ to $z\simeq8$, at $\phi_{\rm LRD} \simeq 6 \times 10^{-5}\,\mathrm{Mpc}^{-3}$~\citep[e.g.,][]{Inayoshi2025First,Kocevski2025,Kokorev2024}. 
This corresponds to approximately $480$ LRDs within the Shin-Uchuu simulation volume of $140^{3}\,h^{-3}\,\mathrm{Mpc}^{3}$.
These values guide our LRD tagging strategy. 

\subsection{LRD formation mechanisms and tagging strategies}\label{sec:mech}

\noindent\textbf{DCBH scenario:}
A possible pathway to produce the massive BHs and the weak or absent stellar components inferred for LRDs within the $\LCDM$ framework is the formation of heavy seeds via DCBHs~\citep[e.g.,][]{Jeon2025,Cenci2025,Pacucci2026}, followed by sustained accretion at super-Eddington rates in the quasi-star phase~\citep[e.g.,][]{Begelman2008}. In this scenario, gas collapses in rare atomic-cooling haloes and forms a massive BH seed of mass $M_{\rm BH} \sim 10^{4-6}\Msun$. 
The formation conditions of a DCBH can be characterized as follows:
(i) the host halo must have a virial temperature $T_{\rm vir} > 10^{4}\,\K$, such that cooling proceeds primarily through atomic hydrogen;
(ii) the gas must remain metal-free in order to suppress gas fragmentation driven by metal-line and dust cooling;
(iii) the halo must be exposed to a sufficiently strong Lyman-Werner (LW) radiation field to dissociate molecular hydrogen, since otherwise $\mathrm{H}_{2}$ cooling would reduce the gas temperature to $\sim200\,\K$, lower the Jeans mass, and induce fragmentation and star formation rather than monolithic collapse~\citep[][]{Bromm2003,Shang2010,Omukai2008,Inayoshi2014,Regan2014,Wise2019}.

Guided by these criteria, we adopt a simplified toy model to identify potential DCBH host haloes in the Shin-Uchuu simulation. We consider all haloes with a virial temperature of $T_{\rm vir} > 10^{4}\K$, classify a halo as pristine if it hosts no galaxy in the associated UniverseMachine catalogue, and require the total LW intensity, $J_{\rm LW} = J_{\rm loc} + J_{\rm bkg}$, to be higher than a critical value\footnote{We note that additional dynamical heating from halo mergers is not considered in this work. Such heating may further enhance the efficiency of DCBH formation in overdense regions~\citep[][]{Li2021,Lupi2021}.}, $J_{\rm crit}$, where $J_{\rm loc}$ represents the contribution from nearby galaxies and $J_{\rm bkg}$ denotes the cosmological background\footnote{Although the cosmological LW background is generally subdominant to the local contribution~\citep[e.g.,][]{Greif2006,Jeon2025}, we include it for completeness.}. 
Following \citet[][]{Greif2006} and \citet[][]{Dijkstra2008}, we approximate
\begin{align}
J_{\rm bkg} &= 10^{2-z/5}, \\
J_{\rm loc} &= \frac{1}{4\pi}\sum_i f_{\rm es}\frac{L_{{\rm LW},i}({\rm SFR}_i)}{4\pi r_i^2},
\end{align}
where the index $i$ runs over nearby galaxies within $100\pkpc$, $r_i$ denotes their distances to the target halo, and $f_{\rm es}=0.5$ is the escape fraction of LW photons~\citep[][]{Agarwal2012,Dijkstra2014}. 
Here $J$ is expressed in units of $10^{-21}\,\rm erg\,s^{-1}\,\cm^{-2}\,Hz^{-1}\,sr^{-1}$. 
We neglect attenuation of the local LW field by the IGM, as LW photons propagate largely unattenuated through the neutral IGM and are absorbed primarily after cosmological redshifting into Lyman-series resonance lines. The corresponding attenuation length is therefore several physical Mpc, substantially exceeding the characteristic separations considered here~\citep[e.g.,][]{Haiman1997,Ahn2009}. This approximation is further supported by the fact that the selected DCBH candidates are typically illuminated by UV-bright companions within $\sim10\pkpc$. Attenuation within the ISM is instead incorporated into the escape fraction $f_{\rm es}$, whose adopted value is broadly consistent with radiation-hydrodynamic simulations, although it remains uncertain~\citep[e.g.,][]{Schauer2017}.

For the LW luminosity of neighboring galaxies, we adopt an empirical UV-to-SFR conversion~\citep[][]{Kennicutt1998} and approximate the LW specific luminosity as
\begin{equation}
L_{\rm LW}\sim 8\times10^{27}\,\left(\frac{\rm SFR}{\Msun/\rm yr}\right)\, \rm erg\,s^{-1}\,Hz^{-1},
\end{equation}
where $\rm SFR$ denotes the star formation rate provided by the UniverseMachine catalogue. 
We set $J_{\rm crit}=1000$~\citep{Sugimura2014}. 
To account for metal pollution driven by galactic winds, we cease DCBH seeding at redshifts below $z=10$~\citep[][]{Dijkstra2014}. 
For any halo that satisfies the above criteria and maintains $J_{\rm LW}>J_{\rm crit}$ for at least one dynamical time (approximated by one tenth of instantaneous Hubble time, $0.1 t_{\rm H}$), we classify it as a potential DCBH host. 
This procedure yields 4,452 candidates at $z=5$. We caution against over-interpreting the predicted abundance, as our toy model is subject to several uncertainties, including the duty cycle of LRDs, the LW intensity estimates, external metal contamination, and the contribution from Pop III stars. To match the observed abundance of LRDs, we randomly down-sample the candidates to 10\%.

\bigskip
\noindent \textbf{SIDM scenario:}
LRDs are believed to have SMBHs that are overmassive relative to their host galaxies compared to local scaling relations \citep{Matthee2024,Maiolino2024}, and some studies even argue that some LRDs may lack substantial host galaxies altogether \citep[e.g.,][]{Chen2025,Chen2025b}. 
This motivates scenarios in which BH seeding has a non-baryonic origin. 
\citet{Jiang2025} explored the feasibility of seeding the massive BHs inferred for LRDs through gravothermal core collapse of SIDM haloes. 
In this framework, LRDs are expected to reside in highly concentrated haloes that form sufficiently early, which enables core collapse to proceed rapidly and be completed by the epoch when LRDs are observed. 
More specifically, the requirement for a halo to seed a massive BH reduces to the condition that the gravothermal core-collapse timescale, $t_{\rm cc}$, be shorter than the time elapsed since halo formation, $t_{\rm form}$.

The core collapse timescale $t_{\rm cc}$ is approximated as~\citep[e.g.,][]{Essig2019}
\begin{equation}\label{eq:tcc}
t_{\rm cc}\approx\frac{150}{C}\frac{1}{\sigma_{\rm eff}\rho_{\rm s}r_{\rm s}}\frac{1}{\sqrt{4\pi G\rho_{\rm s}}},
\end{equation}
where $C=0.75$ is an empirical constant calibrated using $N$-body simulations~\citep[e.g.,][]{Nishikawa2020}, and $\rho_{\rm s}$ and $r_{\rm s}$ denote the scale density and scale radius of the NFW profile followed by the halo at its formation time. 
The factor $\sigma_{\rm eff}$ denotes the effective cross section per particle mass of the self-interactions. 
We follow \citet[][]{Yang2022,Yang2023} to define it as the integral of the velocity-dependent cross section over the velocity distribution of a halo, via 
\begin{equation}\label{eq:sigmaeff}
    \sigma_{\mathrm{eff}}=\frac{1}{2}\int\tilde{v}^{2}\mathrm{d}\tilde{v}\sin^{2}\theta\mathrm{d}\cos\theta \frac{\sigma_0\omega^4}{2[\omega^2+v^2\sin^2(\theta/2)]^2}\tilde{v}^{5}e^{-\tilde{v}^{2}},
\end{equation}
where $\tilde{v} \equiv v/(2v_{\rm eff} )$, and $v_{\rm eff}$ is a characteristic one-dimensional velocity dispersion. 
For an isotropic velocity field, it can be approximated by $v_{\rm eff}\approx v_{\rm max}/\sqrt{3}$, with $v_{\rm max}$ being the maximum circular velocity of the halo.
For the differential cross section, we consider elastic Rutherford scattering through a Yukawa potential where $\sigma_0$ is the normalization at low-velocity end, and $\omega$ is the velocity scale above which the cross section decreases sharply. 

To evaluate the core-collapse time, we traverse the halo merger trees to obtain the halo mass $\Mv$, scale radius $r_{\rm s}$, and concentration $c$ at each redshift, and compute $\rho_{\rm s}$ from Eq.~\ref{eq:rhos}. We adopt a cross section of $\sigma_{\rm 0} = 10 \rm \cm^2/g$ and $\omega=200\, \rm km/s$, which are broadly consistent with constraints from the rotation curves of nearby galaxies~\citep[e.g.,][]{Jia2026}, and use these parameters to derive the effective cross section $\sigma_{\rm eff}$.

However, the numerical resolution of the Shin-Uchuu simulation is just marginally sufficient for probing this scenario.
Since both the core-collapse time, $t_{\rm cc}$, and the formation time, $t_{\rm form}$, depend on halo mass, there exists a characteristic halo mass where BH seeding is most efficient, $\Mv\sim10^8 h^{-1}\Msun$, for the cross section we consider.
This corresponds to roughly 100 DM particles in the Shin-Uchuu simulation.
The marginal numerical resolution impacts the accuracy of concentration measurements \citep{Wang2024}. 
To ensure robustness against such uncertainties, we require the condition, $t_{\rm cc} < t_{\rm form}$, to be satisfied continuously for at least one dynamical time, $t_{\rm dyn}\sim0.1 t_{\rm H}$, before identifying a halo as having undergone core collapse and seeded a BH. 
This procedure yields $437$ LRD candidates, primarily seeded over the redshift range $z\simeq8-18$.

\bigskip
\noindent \textbf{PBH scenario:}
Primordial black holes (PBHs), if they exist, may provide another non-baryonic origin for LRDs \citep[e.g.,][]{Liu2024,Dayal2024,ZhangSY2025,ZhangBorui2025}. PBHs can form from sufficiently large primordial inhomogeneities, either seeded in the initial conditions of the Universe or generated dynamically in the early Universe, and they may span a wide range of masses.

Two classes of scenarios have been discussed in the context of LRDs. In the first, PBHs are born massive, with $M_{\rm PBH}\gtrsim 10^{6}\Msun$~\citep[e.g.,][]{ZhangSY2025}. In this case, existing bounds require a small PBH contribution to the total dark matter budget, $f_{\rm PBH}\lesssim 10^{-4}$~\citep[e.g.,][]{Carr2021,Carr2022}. The low number density implies that the PBH two-point correlation function approaches a white noise plateau on comoving scales below $L_{\rm w}\sim (f_{\rm PBH}\,\rho_{c}/M_{\rm PBH})^{-1/3}\sim 0.1-1\,{\rm cMpc}$. On larger scales, once cosmic structure formation becomes nonlinear, PBHs are expected to trace the underlying matter clustering, $\xi_{mm}$~\citep[][]{Desjacques18,Luca2020}. We note that, if one assumes PBHs are created from the high-$\sigma$ peaks of nearly Gaussian fluctuations, constraints from second-order tensor perturbations and $\mu$-distortions in CMB rule out such a massive PBH scenario~\citep[e.g.,][]{Carr1993,Carr2021,Chluba2012,Chluba2021,Kohri2014}.

In the second, more complex class of scenarios, PBHs form initially at $M_{\rm PBH}\lesssim 10^{2}\Msun$ with strong small-scale clustering in the early Universe and subsequently undergo runaway mergers that produce a population of larger PBHs before Hawking evaporation becomes important~\citep[e.g.,][]{Holst2025,ZhangBorui2025}. This scenario evades the CMB $\mu$-distortion constraints discussed above. The characteristic mass and number density of the final merged PBHs are comparable to those in the first scenario. Therefore, for simplicity, we adopt the same large-scale clustering approximation, while neglecting additional small-scale structure imprinted by the early time clustering and merger history. Finally, PBH dynamics within DM haloes can further modify the one-halo term, so we truncate our clustering prediction on scales around and below the white noise scale $L_{\rm w}$.

\bigskip
\noindent \textbf{Low-spin scenario:} 
\citet{Pacucci2025} interpret LRDs as compact galaxies with exceptionally small effective radii ($R_{\rm eff}\sim 50-300\,\pc$), which they attribute to low halo spin following a simplified version of the model by \citet{Mo1998}, in which galaxy size scales with the spin and virial radius of the host halo, $\Reff \propto \lambda \Rv$.

In this scenario, LRDs correspond to the lowest 1\% tail of the halo spin distribution. 
The redshift evolution of the LRD abundance is governed by cosmological surface-brightness dimming and by the evolution of the critical density, which affects halo virial radii: at higher redshift, compact systems are more common but increasingly more difficult to detect, whereas at lower redshift they are brighter but rarer. 

Guided by this picture, we use the UniverseMachine mock galaxy catalogue to select candidate LRD hosts as the lowest 1\% in halo spin among haloes hosting UV-bright galaxies with $M_{\rm UV}<-17$.
The UV brightness threshold follows \citealt{Kocevski2025} and the practice of \citealt{Pacucci2025}. 
For subhaloes, we adopt the halo spin evaluated at the epoch of first infall, assuming that the size of a satellite galaxy is set by the spin and virial radius of its host halo when it was still a distinct halo.
This selection yields a total of 850 objects within our simulation volume.

\subsection{Clustering of LRDs}\label{sec:clstr}

\subsubsection{Measurements}
We now evaluate the auto-correlation functions of LRD host haloes and normal galaxies, as well as their cross-correlation functions. 
Here, normal galaxies are defined as UniverseMachine galaxies with $M_{\rm UV}<-17$. 
The method used to measure the auto-correlation function is described in \se{2pcf}. 
The cross-correlation function is defined as
\begin{equation}
    \xi_{12}(r)=\frac{D_1D_2(r)-D_1R_2(r)-D_2R_1(r)+R_1R_2(r)}{R_1R_2(r)},
\end{equation}
where $D_1D_2(r)$ is the number of cross pairs between the two data catalogues (i.e., LRDs and normal galaxies), while $D_1R_2(r)$, $D_2R_1(r)$, and $R_1R_2(r)$ denote the corresponding pair counts involving random catalogues.

To facilitate comparison with observations, we compute the projected correlation function, defined as
\begin{equation}
    w_p(r_p)=2\times\int_{0}^{r_{\pi,{\rm max}}}\xi(r_p,r_\pi)\,{\rm d}\,r_\pi,
\end{equation}
where $r_p$ and $r_\pi$ are the transverse and line-of-sight separations, respectively. 
We adopt $r_{\pi,{\rm max}}=8\,h^{-1}\rm Mpc$, corresponding to a velocity difference of approximately $1000~\mathrm{km\,s^{-1}}$ at $z\simeq 4-6$, which effectively captures the clustering signal from large-scale structure. 

\begin{figure*}
    \centering
    \includegraphics[width=0.8\linewidth]{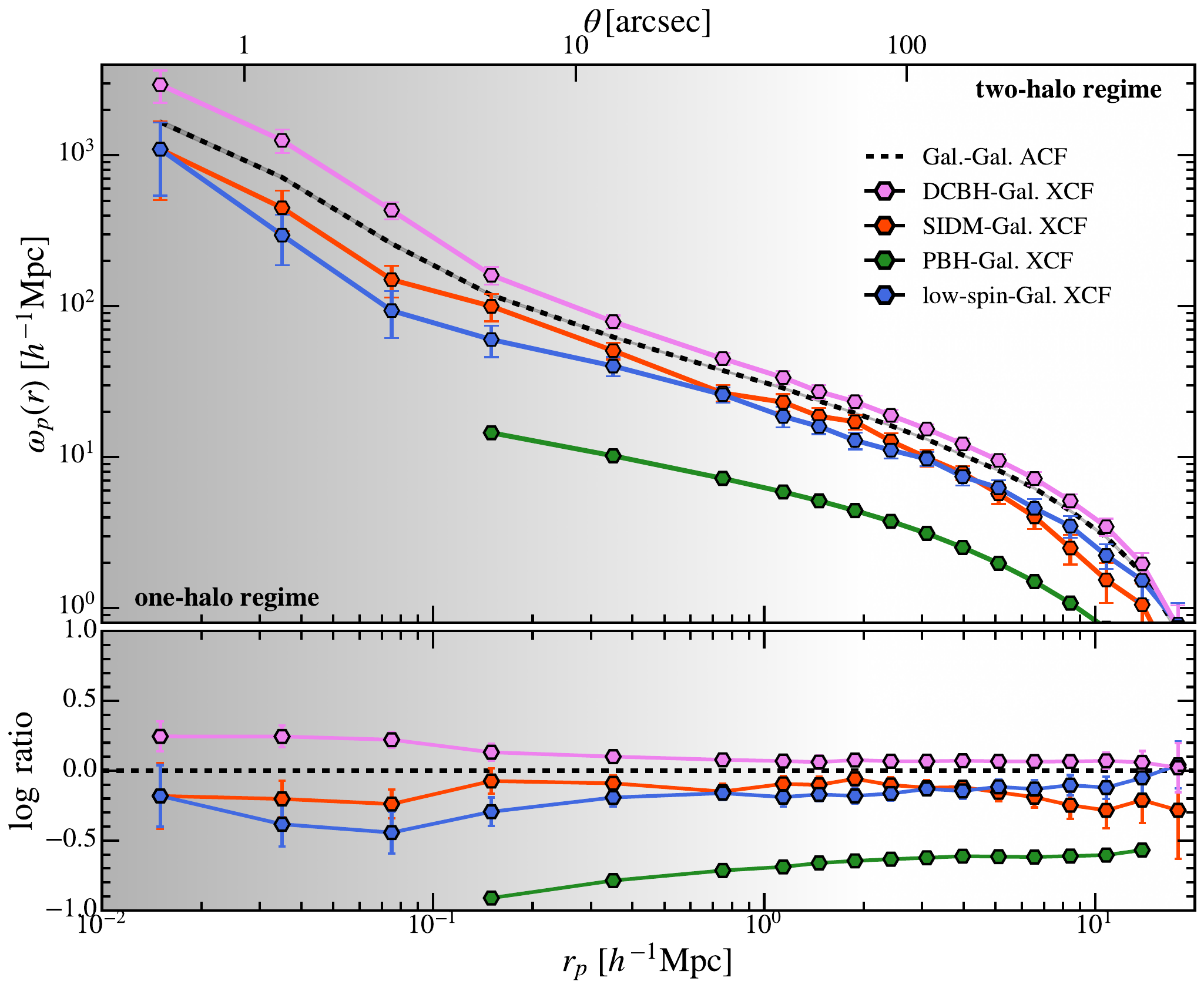}
    \caption{Projected correlation functions between galaxies with $M_{\rm UV}<-17$ and LRDs at $z=5$. 
    The black curve shows the auto-correlation function (ACF) of the normal galaxy sample, while colored curves denote the galaxy-LRD  cross-correlation functions (XCFs) for the SIDM, DCBH, low-spin, and PBH scenarios. 
    Shaded regions and error bars indicate uncertainties estimated from 1000 bootstrap realizations, with the DCBH case additionally resampled according to the down-sampling factor. 
    For the PBH scenario, the two-halo term is computed assuming PBHs trace the matter distribution. 
    The top axis shows the corresponding angular separation at $z=5$. 
    The lower panel displays the ratio of the LRD-galaxy cross-correlation to the galaxy auto-correlation. 
    \quad
    Among the four scenarios, the DCBH case exhibits stronger clustering than the galaxy-galaxy ACF on all scales.  
    The low-spin case yields systematically weaker clustering across all scales. 
    The SIDM case also shows weaker clustering on all scales compared to the galaxy-galaxy ACF, but exhibits a mild enhancement on small scales relative to the low-spin case.
    }
    \label{fig:lrd_clst}
\end{figure*}

\Fig{lrd_clst} presents the projected correlation functions, where the black curves show the auto-correlation function of the normal galaxy sample, while the colored curves denote the galaxy-LRD cross-correlation functions for the four LRD-selection scenarios, as labeled. 
Uncertainties for the SIDM and low-spin scenarios are estimated via bootstrap resampling with replacement. 
For the DCBH scenario, we resample the candidate catalogue according to the downsampling factor in each realization, and repeat this procedure 1000 times to estimate the uncertainties. 
For the PBH scenario, we assume that PBHs trace the matter distribution on large scales and compute the galaxy-PBH cross-correlation in the two-halo regime using $\xi_{\rm mm}(r,z)$. 
We evaluate $\xi_{\rm mm}$ with \textsc{halomod}\footnote{\url{https://halomod.readthedocs.io/en/latest/index.html}}, adopting the halo-bias calibration of \citet{Tinker2010}, the transfer function from \textsc{camb}\footnote{\url{https://camb.readthedocs.io/en/latest/}}, and the linear growth factor from \citet{Challinor2011}. 
The upper axis indicates the corresponding angular separations at $z=5$, converted from comoving distances. The bottom panel of \Fig{lrd_clst} shows the ratio between the LRD-galaxy cross-correlation function and the galaxy auto-correlation function.

The clustering strength of the DCBH model is the strongest across most scales, followed by the SIDM and low-spin models. At large scales, the low-spin model shows slightly stronger clustering than the SIDM model, while in the one-halo regime the SIDM signal becomes stronger. The PBH model exhibits the weakest clustering signal in the two-halo regime, and is truncated by hand at smaller scales.

Some observational studies perform clustering analyses of \textit{JWST}-selected AGN samples by cross-correlating them with specific galaxy populations~\citep{Arita2024,Lin2025,Zhang2026}. However, these galaxy samples differ from the galaxy catalogue adopted in this work and are difficult to reproduce with the UniverseMachine catalogue, which prevents a direct comparison. Nevertheless, the dependence on the chosen galaxy sample can be removed by subtracting the galaxy auto-correlation, yielding a clean estimate of the bias factor, which we compare in \Fig{bias}.

\subsubsection{Large-scale clustering of LRDs}\label{sec:ls_clstr}
To interpret the results, we first quantify the clustering strength in the two-halo term regime.
Following \citet{Arita2024}, we estimate the linear bias parameter $b$ as the ratio between the clustering amplitude of a given tracer and that of the underlying DM, evaluated at a fixed comoving scale of $8\,h^{-1}\mathrm{Mpc}$,
\begin{equation}
b=\sqrt{\frac{\xi(8,z)}{\xi_{\rm mm}(8,z)}},
\end{equation}
where $\xi(8,z)$ is obtained by fitting the real-space auto- or cross-correlation function with a power-law form over the range $r=1-10\,h^{-1}\mathrm{Mpc}$ and evaluating the best-fit model at $8\,h^{-1}\mathrm{Mpc}$. 
We estimate the linear bias parameter for the galaxy-galaxy auto-correlation function, $b_{\rm gal}$, and for the galaxy-LRD cross-correlation function, $b_{\rm gL}$. 
Under the assumption of linear bias on these scales, the cross-correlation amplitude satisfies 
\begin{equation}
    b_{\rm LRD} \simeq b_{\rm gL}^2/\, b_{\rm gal},
\end{equation}
which allows us to infer the bias of the LRD population, $b_{\rm LRD}$~\citep[][]{Mountrichas2011}.

\begin{figure}
    \centering
    \includegraphics[width=0.95\linewidth]{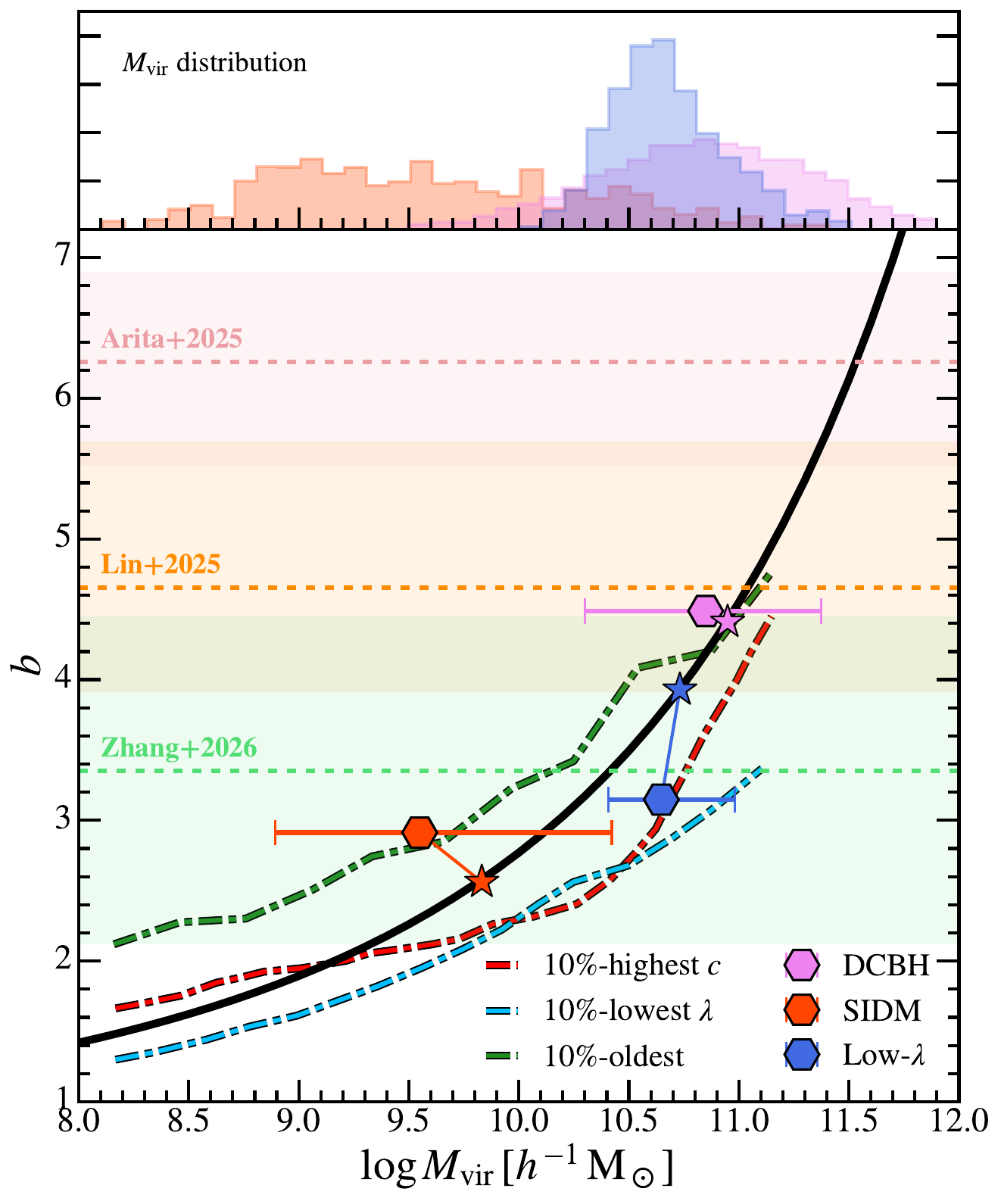}
    \caption{
    Large-scale linear bias inferred for LRD hosts under different formation scenarios observed at $z\simeq5$. 
    The main panel shows the measured bias $b$ as a function of host halo virial mass, for the DCBH, SIDM, and low-spin scenarios, with symbols indicating the median and horizontal bars denoting the $1\sigma$ scatter of the corresponding host-halo mass distributions. 
    The solid black curve shows the standard halo bias-mass relation from \citet{Tinker2010}, while the dotted curves represent the bias-mass relations for haloes of the highest 10\% concentration, lowest 10\% spin, and earliest 10\% formation time, as labeled. 
    For each scenario, the star marker denotes the effective bias $b_{\rm eff}$ expected from halo-mass selection alone, computed by convolving the host-halo mass distribution with the standard halo bias relation. 
    Shaded horizontal bands indicate the halo bias ranges inferred from clustering analyses of spectroscopically confirmed \textit{JWST}-selected AGN samples, including both LRD-like and unobscured populations~\citep[e.g.,][]{Lin2025,Arita2024}, and from the recent photometrically selected LRD sample of \citet{Zhang2026}.
    The upper panel shows the normalized host-halo mass distributions for each scenario. 
    Differences in the inferred bias reflect the combined effects of halo mass and secondary bias.
    \quad
    The DCBH scenario yields a relatively high bias of $b\ga 4$. 
    The low-spin and SIDM scenarios produce similar biases of $b\sim3$, but correspond to rather different halo mass ranges. 
    }
    \label{fig:bias}
\end{figure}

\Fig{bias} presents the inferred large-scale linear bias for different LRD-host scenarios, together with their corresponding host-halo mass distributions at $z\simeq5$. 
These measurements are compared with the standard halo bias-mass relation from \citet{Tinker2010}, as well as with the bias-mass relations considering haloes of the highest 10\% concentration, lowest 10\% spin, and earliest 10\% formation time.

Since the different scenarios naturally select LRD hosts with different halo-mass distributions, as illustrated in the upper panel of \Fig{bias}, part of the variation in clustering strength can be attributed trivially to mass differences, rather than to assembly bias specific to each scenario. 
To quantify this mass-driven contribution, we also compute, for each scenario, a mass-only ``effective'' bias, $b_{\rm eff}$, by convolving the scenario-specific mass distribution with the standard \citealt{Tinker2010} bias-mass relation. 
This mass-only baseline provides a reference basis against which deviations can be attributed to secondary biases.

For comparison with observations, we overlay host-halo mass estimates from clustering analyses of spectroscopically confirmed \textit{JWST} AGN samples, including both LRD-like and unobscured populations \citep{Arita2024,Lin2025}. We also include the recent clustering measurement of a more stringently selected LRD sample by \citet{Zhang2026}, although the latter is based primarily on photometric redshifts and is therefore subject to larger redshift uncertainties.

The DCBH scenario predicts the strongest clustering, with a bias of $b_{\rm LRD} \simeq 4.50$, reaching the regime of the observed AGN samples~\citep{Lin2025}. 
Slightly enhanced clustering relative to the standard bias-mass relation likely arises because the LW radiation criterion preferentially selects haloes in overdense environments. 
The resulting population spans a relatively broad halo-mass range of $\Mv\ga 10^{9.5}\,h^{-1}\Msun$, with a median of $\Mv\sim10^{10.8}h^{-1}\Msun$. 

The low-spin scenario gives a bias of $b \simeq 3.16$, which is lower than the standard bias for haloes of comparable masses, driven primarily by the secondary bias associated with halo spin. 
The halo mass distribution is relatively narrow and massive, peaked around $\Mv\sim 10^{10.6}h^{-1}\Msun$.

The SIDM scenario likewise yields a weak clustering signal, with $b_{\rm LRD} \simeq 2.92$. 
However, this is primarily driven by its host haloes spanning a broad but systematically lower mass distribution. 
The halo mass at the time of core-collapse is typically $\Mv\sim10^{8-9}h^{-1}\Msun$, after which the systems can continue to grow to higher masses. Both the low-spin and SIDM scenarios predict large-scale bias factors that fall within the current observational uncertainty reported by \citet{Zhang2026}, indicating that the present clustering measurements alone are not yet sufficient to distinguish between them.

For the PBH scenario, assuming that PBHs trace the underlying matter distribution on large scales and that they become LRDs stochastically, the expected large-scale bias is $b_{\rm LRD} = 1$ by definition.
The halo-mass distribution depends on theoretical details such as the initial PBH mass spectrum and lies beyond the scope of this work.

\subsubsection{Small-scale clustering and satellite fraction}\label{sec:ss_clstr}

\begin{figure}
    \centering
    \includegraphics[width=0.95\linewidth]{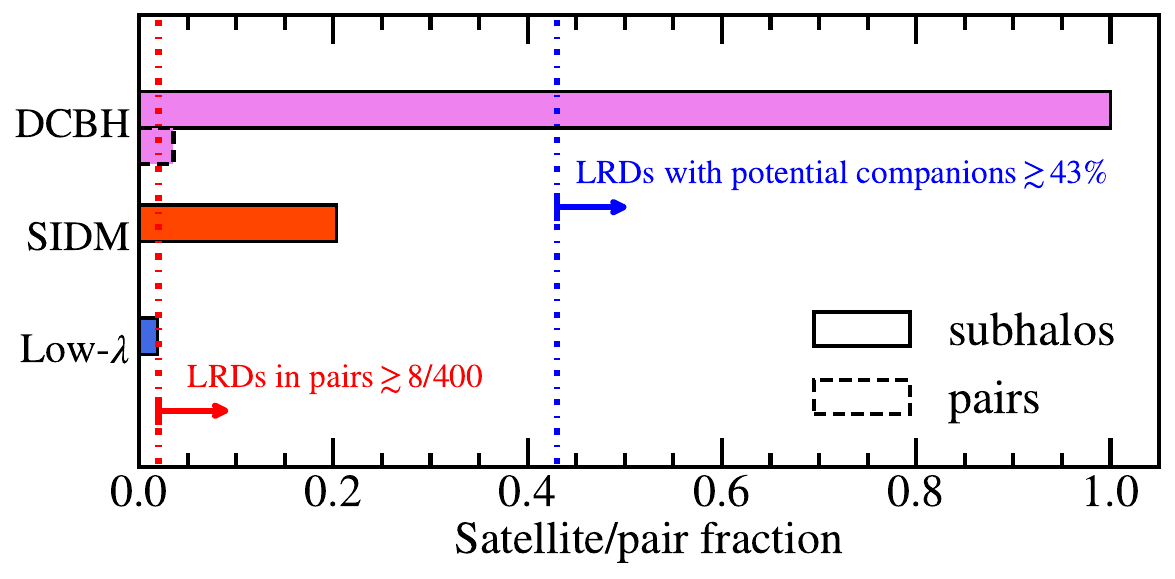}
    \caption{Fraction of LRDs associated with subhaloes and close pairs, predicted by the different formation scenarios. For comparison, we adopt the fraction of LRDs exhibiting spatially offset UV emission from \citet{Baggen2026}, which could be suggestive of companion galaxies. We also compile close-pair statistics from \citet{Tanaka2024,Merida2025}. The DCBH scenario, by construction, places LRDs around massive centrals and therefore produces a higher fraction of close pairs, while the SIDM model, owing to its lower halo mass range, predicts a modest satellite fraction.
    }
    \label{fig:sat_frac}
\end{figure}

An important observational feature of LRDs emerging from preliminary clustering analyses is their enhanced small-scale clustering~\citep[e.g.,][]{Zhuang2025}, a high incidence of offset companions\footnote{While companions of LRDs have been commonly reported, their nature remains uncertain. Some resolved systems exhibit clear galaxy-like features~\citep[e.g.,][]{Baggen2026}, whereas in other cases the emission appears to arise from nebular gas associated with the LRD itself~\citep[e.g.,][]{Chen2025b}.}~\citep[e.g.,][]{Rinaldi2025,Baggen2026,Chen2025b}, and a significant pair fraction~\citep[e.g.,][]{Tanaka2024,Merida2025}. Here, offset companions refer to UV-emitting sources displaced by a few kiloparsecs from the optical emission of an LRD. Notably, four close LRD pairs (dual-LRD systems) have been identified from photometric observations~\citep[][]{Tanaka2024,Merida2025} among a total sample of $\sim400$ LRDs, with additional candidates likely unresolved. Motivated by these observations, we quantify both the abundance of satellite LRDs and the incidence of close LRD pairs in the different formation scenarios.


As already shown in \Fig{lrd_clst}, the DCBH scenario predicts the strongest small-scale clustering relative to normal galaxies on scales of 10-100$h^{-1}\kpc$, followed by the SIDM and low-spin scenarios.

At even smaller spatial scales, where pairs and close companions become relevant, subhalo statistics provide a useful complement to the clustering analysis. 
One notable trend from the subhalo mass function is that the abundance of satellites decreases dramatically with increasing subhalo mass. In this sense, the abundance of satellite LRD hosts is $\sim3$ dex higher in haloes with masses of order $10^{8-9}\,h^{-1}\Msun$ than in haloes with masses near $10^{11}\,h^{-1}\Msun$.
Indeed, low-mass haloes of order $10^{8-9}\,h^{-1}\Msun$ are also more naturally consistent with the lack of prominent stellar components in LRDs, since haloes with masses of $\sim10^{11}\,h^{-1}\Msun$ are expected to host substantial and spatially extended stellar components at these epochs.
In most cosmological hydrodynamical simulations capable of resolving systems with $\Mv\sim10^{11}\,h^{-1}\Msun$, galaxy sizes are generally of order $\kpc$~\citep[e.g.,][]{Shen2024,McClymont2025}. 

To this end, we further investigate the predicted abundances of satellite LRDs and close LRD pairs under the different scenarios. 
In \Fig{sat_frac}, we show the fraction of satellite LRDs and that of LRDs in pairs.
To estimate the pair fraction, we randomly draw 480 LRDs from each scenario, or use the full sample when fewer are available, and repeat this procedure 1000 times, counting the number of systems that satisfy the pair criteria in each realization.
Here, a satellite LRD is defined as an LRD associated with a subhalo, either at the observed epoch or at any earlier time prior to subhalo disruption. 
An LRD pair is defined as two (or more) LRDs that satisfy at least one of the following criteria: (i) their host haloes have undergone a merger; (ii) their hosting subhaloes reside within the same host halo; or (iii) their projected separation at the observed epoch is smaller than $5\,\mathrm{pkpc}$. The results are further compared with compiled observations~\citep[e.g.,][]{Tanaka2024,Merida2025,Baggen2026}.

The DCBH scenario most efficiently produces LRD satellites and pairs.
This is, to some extent, by construction, since DCBH hosts preferentially form as satellites of UV-bright central galaxies, allowing multiple LRDs to arise within the same halo. 

The SIDM scenario yields the second-highest satellite abundance, primarily because it preferentially selects lower-mass host haloes, which have a higher probability of being accreted as subhaloes. 
However, no resolved LRD pairs are found in this case, likely because high-concentration haloes at the relevant masses exhibit relatively weak clustering at the seeding epoch. 

The low-spin scenario produces the lowest number of satellite LRDs and likewise yields no LRD pairs.

\subsection{Evaluation and limitations of the results}\label{sec:limit}
It is important to clarify some uncertainties in our results, which mainly arise from the tagging criteria adopted for LRD hosts. 

For the DCBH scenario, our implementation only captures DCBH formation as a channel linked to close halo pairs~\citep[e.g.,][]{Dijkstra2008,Visbal2014}, while alternative pathways, such as DCBH formation following prior first star formation within the same halo~\citep[e.g.,][]{Susa2007}, are not considered. The inferred host halo mass distribution does not directly reflect the subhalo masses at which DCBH seeds initially form. The majority of these seed-hosting subhaloes are fully disrupted by the observed epoch, such that the merger trees point to their surviving distinct host haloes. This behavior is a natural outcome of hierarchical structure formation rather than a numerical artifact. Instead, it highlights that DCBH formation is preferentially associated with massive central haloes that can both provide sufficiently strong LW radiation fields and host intermediate-mass atomic-cooling subhaloes with $T_{\rm vir}>10^{4}\,\mathrm{K}$. As a result, the large-scale clustering of DCBHs primarily traces the clustering of their central host haloes.

In the one-halo regime, the enhanced clustering signal also depends sensitively on the treatment of the LW radiation field and the assumed critical intensity $J_{\rm crit}$. Previous studies have shown that adopting a more stringent LW threshold can substantially amplify the small-scale clustering signal, with secondary effects that also propagate to larger scales~\citep[e.g.,][]{Agarwal2012}. In addition, external metal enrichment from neighboring galaxies and Pop-III stars is not modeled here, which would further suppress the number of viable DCBH hosts and likely weaken the small-scale clustering~\citep[e.g.][]{Dijkstra2014}.

In the SIDM scenario, first, we note that the choice of the self-interaction cross section does not qualitatively affect our conclusions, although it does influence the characteristic halo mass range over which core collapse and BH seeding occur ~\citep{Jiang2025}. 
Specifically, lower $\sigma_0$ and higher $\omega$ shift the characteristic halo mass for BH seeding toward higher values.
The parameters adopted here are chosen such that core collapse can occur in haloes resolved in the simulation, with $\Mv\ga 10^{8}h^{-1}\Msun$. 
Even at this cross section, we can explicitly identify seed hosts only above this mass, while a non-negligible fraction of BHs may form in lower-mass, unresolved haloes. 
The inferred host-halo mass distribution should therefore be regarded as incomplete and missing a lower-mass tail. 
If lower-mass hosts were included, the true large-scale clustering of SIDM-seeded LRDs would likely be even weaker, although this could be partially offset by a higher satellite fraction and an increased incidence of close pairs.

Second, our analysis assumes that SIDM does not significantly modify the formation of large-scale structure or the overall concentration distribution, and instead adopts concentration measurements from the CDM simulations. 
This assumption is supported by both physical considerations and existing numerical results, as SIDM effects become pronounced primarily within radii $r\lesssim r_{\rm s}$~\citep[][]{Yang2024,Rocha2013}. 
On small scales, however, recent simulations~\citep[e.g.,][]{Romanello2025} have shown that SIDM can suppress the one-halo term of the clustering signal, which may reduce the small-scale clustering amplitude relative to our predictions. 

Overall, we can safely conclude that the SIDM scenario features the lowest halo-mass distribution among the different scenarios considered here. 
Interestingly, the secondary bias associated with early formation and high concentration largely cancels out.

For the low-spin scenario, the reduced clustering relative to the standard mass-based bias is a direct consequence of the secondary bias associated with halo spin. This qualitative behavior is robust and largely insensitive to the detailed host halo mass distribution. 

In addition, halo spin may be physically connected to DCBH formation, although its role remains debated. 
On the one hand, some studies suggest that low angular momentum more easily avoids centrifugal support and facilitates large-scale inflows, allowing a larger fraction of the gas to collapse into a DCBH and subsequently grow~\citep[e.g.,][]{Lodato2006,Agarwal2012,Suazo2019,Bhowmick2021}. On the other hand, high angular momentum may also accelerate core collapse through the so-called “gravo-gyro” instability~\citep[e.g.,][]{Hachisu1979,Dekel2025}. Other studies, however, argue that angular momentum is not a critical factor in the collapse process of DCBH-forming haloes~\citep[e.g.,][]{Chon2016,Mone2025}, partly because the angular momentum of the gas does not necessarily trace that of the dark matter halo, especially on small scales where the dynamics of the nuclear disc dominate \citep{Dubois2012,Bonoli2014,Danovich2015}.
Regardless of the exact mechanism, if DCBH formation preferentially occurs in haloes with certain spin properties, the resulting population would inherit the clustering signal associated with halo spin bias.

Finally, we note that the discussion above is based on the assumption that LRDs arise from a single formation channel. 
In reality, LRDs may originate from multiple channels and therefore exhibit a superposition of different clustering signatures. 
In addition, uncertainties related to the subsequent dynamics of host haloes and BH seeds, including tidal disruption of subhaloes and wandering BHs, are not considered here. 
Nevertheless, our results still provide useful constraints on the relative contributions of these channels and offer a framework for interpreting future clustering measurements.

\section{Conclusions}\label{sec:conclusions}
In this work, we provide the first systematic characterization of halo assembly bias in the early Universe, examining its dependence on formation time, concentration, and spin across redshifts, using the Shin-Uchuu simulation. 
We then explore how secondary bias can be leveraged in the \textit{JWST} era, using LRDs as a case study to predict their clustering signatures and host-halo mass ranges under different formation scenarios. 
Our main results are summarized below.

\begin{itemize}
\item 
At all redshifts, earlier-forming haloes are more strongly clustered than later-forming haloes at fixed mass. 
High-concentration haloes show enhanced clustering at the low-mass end, whereas this trend reverses toward higher masses. 
Low-spin haloes show reduced clustering across the full mass range at $z\ga 4$.
Toward intermediate and lower redshifts, the clustering signal of low-spin haloes increases progressively at the low-mass end.

\item 
There are clear sign reversals in both concentration-related and spin-related bias signals, and the critical mass at which the assembly-bias signal flips sign depends on redshift. 
For concentration, the characteristic crossover mass above which high-concentration haloes become less clustered broadly follows a peak height of $\nu \simeq 1.5$. 
Notably, in contrast to the local Universe, this implies that the majority of galaxies associated with high-concentration haloes at high redshift are expected to exhibit weaker clustering.
For spin, the crossover mass below which low-spin haloes become more strongly clustered occurs near $\nu \simeq 0.75$, but this only holds for $z\la 3$. At higher redshifts, low-spin haloes are always less clustered, regardless of mass. 

\item 
Different proposed LRD formation channels populate systematically different halo-mass regimes, which strongly influences their clustering signatures. 
The DCBH scenario selects relatively massive haloes ($\Mv\sim 10^{10.8}h^{-1}\Msun$) and predicts the strongest large-scale bias ($b\gtrsim4$).
The low-spin scenario occupies similarly massive haloes but exhibits reduced bias ($b\sim3$) due to spin-related secondary bias. 
The SIDM scenario favors substantially lower characteristic halo masses, peaking at $\Mv\sim10^{9-9.5}h^{-1}\Msun$, and exhibits the broadest halo mass distribution, extending down to the lowest masses resolved in the simulation ($\Mv\ga 10^{8}h^{-1}\Msun$). 
Its large-scale clustering is correspondingly weak and overall comparable to that of the low-spin scenario. However, this arises from a combination of effects, namely, the lower halo masses and the partial cancellation of secondary bias associated with early formation and high concentration.  
In the PBH scenario, where seeds trace the matter distribution by construction, no large-scale bias is expected. 
These differences imply that host-halo mass distributions constitute a primary discriminant among LRD formation models.
Despite the halo-mass differences, assembly bias associated with secondary halo properties still plays a non-negligible role in interpreting clustering-based halo masses.

\item 
The different scenarios make distinct predictions for satellite fractions and close LRD pairs. 
The DCBH scenario most naturally produces multiple LRDs within the same massive environment and hence predicts enhanced pair counts. 
The SIDM scenario, despite exhibiting weaker large-scale clustering, can yield a relatively high satellite fraction because it preferentially selects lower-mass haloes, which are more likely to be accreted as satellites than more massive systems.
The low-spin scenario predicts the lowest satellite abundance. 
These small-scale signatures offer an additional observational avenue for distinguishing formation models.

\end{itemize}

Overall, we conclude that halo assembly bias persists into the early Universe, albeit with a reduced amplitude relative to later times. This underscores the importance of accounting for secondary bias when inferring halo masses of high-$z$ galaxies from clustering measurements. Neglecting assembly bias may lead to systematic offsets in halo mass estimates of up to 1 dex.

For LRDs, our study predicts that different formation scenarios lead to distinct clustering signatures and host-halo mass ranges. These predictions can be tested with upcoming clustering measurements. 
If the cross-correlation between LRDs and normal galaxies with $M_{\rm UV} < -17$ on large spatial scales ($\gtrsim 1\,h^{-1}\Mpc$) is stronger than the auto-correlation of the normal galaxy population, the DCBH scenario would be favored. 
If instead the cross-correlation is weaker, the SIDM and low-spin scenarios would be preferred.

These two scenarios can be further distinguished using small-scale clustering measurements and satellite/pair statistics, as the SIDM model predicts stronger small-scale clustering and a higher satellite fraction than the low-spin model. 
We emphasize that the broader and generally lower halo mass range, which extends to values comparable to or only slightly above the atomic-cooling threshold, is a distinctive feature of the SIDM model, making it particularly appealing given that a substantial fraction of LRDs seem to lack prominent galaxy components.

Finally, if the cross-correlation amplitude is an order of magnitude lower than the auto-correlation of normal galaxies, the PBH scenario may be favored. We leave it to forthcoming observational studies to discriminate among these possibilities.


\section*{Acknowledgements}
We thank Liang Gao, Qiao Wang, Huangyu Xiao,  and Wei-Xiang Feng for useful discussions.
FJ acknowledges support by the National Natural Science Foundation of China (NSFC, Grant No. 12473007), China Manned Space Program with Grant No. CMS-CSST-2025-A03, National Key R\&D Program of China (Grant No. 2025YFA1614103).
ZJ acknowledges support by the Beijing Natural Science Foundation (Grant No. QY23018).
LCH was supported by the China Manned Space Program (CMS-CSST-2025-A09) and the National Science Foundation of China (12233001).
KI was supported by the National Natural Science Foundation of China (12573015, W2532003, 1251101148, 12233001, 12473037), the Beijing Natural Science Foundation (IS25003), and the China Manned Space Program (CMS-CSST- 2025-A09).
We acknowledge the High-performance Computing Platform of Peking University for providing computational resources and support.

\section*{Data Availability}

The analysis pipeline and data products are available upon request to the corresponding authors.



\bibliographystyle{mnras}
\bibliography{ref}





\bsp	
\label{lastpage}
\end{document}